\documentclass[aps,prd,showpacs,onecolumn,preprintnumbers,notitlepage,
nofootinbib,amssymb,10pt,amsfont,singlespacing] {revtex4-1}
\usepackage{amsmath,bm}
\usepackage{times}
\usepackage{braket}
\usepackage{color,graphicx}
\usepackage{slashed}
\usepackage{CJK,upgreek,fancyhdr}
\usepackage{hyperref}
\usepackage{multirow}

\definecolor{Red}{rgb}{1.,0.,0.}
\definecolor{Blue}{rgb}{0.,0.,1.}
\definecolor{nicered}{rgb}{0.7,0.1,0.2}
\definecolor{nicegreen}{rgb}{0.1,0.4,0.2}

\hypersetup{colorlinks,citecolor=nicegreen,linkcolor=nicered,urlcolor=magenta,anchorcolor=blue}

\newcommand{\RNum}[1]{\uppercase\expandafter{\romannumeral #1\relax}}

\begin{document}
\title{\boldmath \texorpdfstring{Twist-3 contributions to $\gamma\gamma\rightarrow\pi^+\pi^-,K^+K^-$}{Lg} processes in perturbative QCD approach}

\author{Cong~Wang$^1$}
\email
[Electronic address: ]
{wangj@mails.ccnu.edu.cn}

\author{Jun-Kang~He$^1$}
\email
[Electronic address: ]
{hejk@mails.ccnu.edu.cn}

\author{Ming-Zhen~Zhou$^2$}
\email
[Electronic address: ]
{zhoumz@swu.edu.cn}

\affiliation{$^1$Institute of Particle Physics and Key Laboratory of Quark and Lepton Physics~(MOE),\\
Central China Normal University, Wuhan, Hubei 430079, P.~R.~China\\
$^2$School of Physical Science and Technology, Southwest University, Chongqing 400715, P.~R.~China}



\begin{abstract}
As one of the simplest hadronic processes, $\gamma\gamma\rightarrow M^{+}M^{-}$ ($M=\pi,K$) could be a good testing ground for our understanding of the perturbative and nonperturbative structure of QCD, and will be studied with high precision at BELLE-\RNum{2} in the near future. In this paper, we revisit these processes with twist-3 corrections in the perturbative QCD approach based on the $k_{T}$ factorization theorem, in which transverse degrees of freedom as well as resummation effects are taken into account. The influence of the distribution amplitudes on the cross sections are discussed in detail. Our work shows that not only the transverse momentum effects but also the twist-3 corrections play a significant role in the processes $\gamma\gamma\rightarrow M^{+}M^{-}$ in the intermediate energy region. Especially in the few GeV region, the twist-3 contributions become dominant in the cross sections. And it is noteworthy that both the twist-3 result of the $\pi^{+}\pi^{-}$ cross section and that of the $K^{+}K^{-}$ cross section agree well with the BELLE and ALEPH measurements. For the pion and kaon angular distributions, there still exist discrepancies between our results and the experimental measurements. Possible reasons for these discrepancies are discussed briefly.
\end{abstract}


\maketitle

\section{Introduction}
A meaningful and historic subject of the perturbative QCD, which has received considerable attention in the past few decades, is the study of the exclusive processes at large momentum transfer. The pioneer works in this area are performed by Efremov and Radyushkin~\cite{Efremov:1978rn} and, independently, Brodsky and Lepage~\cite{Lepage:1980fj}. They pointed out that the exclusive processes at large momentum transfer can be factorized into perturbatively calculable kernels and hadronic wave functions. Although there is general agreement that perturbative QCD is able to make reliable predictions for the exclusive processes in the large energy region~\cite{Chernyak:1983ej}, the applicability of perturbative QCD to these processes in the intermediate energy region has been developed in controversy. In the field of exclusive processes, two-photon collisions $\gamma\gamma\rightarrow\pi^+\pi^-,K^+K^-$~\cite{Brodsky:2000dq,Brodsky:2005wk,Chernyak:2012pw,Chernyak:2014wra,Brodsky:2015nf} are the specially ones with the initial states simple and controllable and the strong interactions only in the final states. These characteristics make the processes $\gamma\gamma\rightarrow\pi^+\pi^-,K^+K^-$ be a good testing ground for our understanding of the perturbative and nonperturbative structure of QCD.

In the perturbative QCD approach based on the collinear factorization, the first investigation of the processes $\gamma\gamma\rightarrow\pi^+\pi^-, K^+K^-$ was carried out by Brodsky and Lepage\cite{Brodsky:1981rp}. At leading-twist leading-order level, they gave detailed calculations of the helicity amplitudes of these processes and presented the known relation between the differential cross section and the timelike electromagnetic form factor:
\begin{eqnarray}\label{relation}
\frac{\mathrm{d}\sigma(\gamma\gamma\rightarrow M^{+}M^{-})}{\mathrm{d}|\cos\theta|}\approx \frac{8\pi\alpha^{2}}{Q^{2}}\frac{|F_{M}(Q^{2})|^{2}}{\sin^{4}\theta}.
\end{eqnarray}
Several years later, Ni\v{z}i\'{c}~\cite{Nizic:1987sw} performed the leading-twist next-to-leading-order perturbative QCD calculations for the these processes. In Refs.~\cite{Nizic:1987sw,Duplancic:2006nv}, it was found that both the leading-order and next-to-leading-order perturbative QCD predictions are almost an order of magnitude smaller than the experimental data~\cite{Aihara:1986qk,Heister:2003ae,Abe:2003vn,Nakazawa:2004gu,Mori:2007bu} in the intermediate energy region. Motivated by the calculation of the pion electromagnetic form factor~\cite{Geshkenbein:1984qn} in which the nonleading twist contributions become the dominant one in a few GeV region, Gorsky~\cite{Gorsky:1989ev} investigated the two-photon process $\gamma\gamma\rightarrow\pi^{+}\pi^{-}$ with the higher-twist corrections. It was pointed out that the two-parton twist-3 contributions
are logarithmically larger than the leading-twist contributions in the intermediate energy region due to the chiral enhancement effects. And this conclusion had also been confirmed in Ref.~\cite{Wang:2015mod} with the BHL prescription~\cite{Lepage:1982gd}. But in a few GeV region, the predicted cross sections with higher-twist corrections~\cite{Gorsky:1989ev,Wang:2015mod} are several times
larger than their experimental measurements~\cite{Aihara:1986qk,Heister:2003ae,Abe:2003vn,Nakazawa:2004gu,Mori:2007bu}.

As it is known that the higher-power effects from the intrinsic transverse momentum play a crucial role in the pion electromagnetic form factor at the scale of few GeV~\cite{Li:1992nu,Jakob:1993iw}, one may expect that the same situation can also be found in the two-photon processes $\gamma\gamma\rightarrow\pi^+\pi^-, K^+K^-$. Based on the $k_{T}$ factorization theorem~\cite{Botts:1989kf,Li:1992nu} in which the transverse momentum dependence was retained, the leading-twist perturbative QCD predictions for the cross sections of these two-photon processes were obtained in Refs.~\cite{Farrar:1989wb,Coriano:1994nh,Coriano:1998ge,Vogt:1999sw,Vogt:2000bz,Hsieh:2004ee}. From the analysis of the differential cross sections obtained in both perturbative QCD and QCD sum rule, the conclusion that the transition from nonperturbative to perturbative QCD in $\gamma\gamma\rightarrow\pi^+\pi^-, K^+K^-$ occurs at $Q\approx2~\mathrm{GeV}$ was drawn in Refs.~\cite{Coriano:1994nh,Coriano:1998ge,Hsieh:2004ee}. However, the numerical results in Refs.~\cite{Vogt:1999sw,Vogt:2000bz} show that the twist-2 cross sections of the $\gamma\gamma\rightarrow\pi^+\pi^-, K^+K^-$ in the $k_{T}$ factorization are still much smaller than the experimental data in the intermediate energy region.

To consider the higher-twist effects for the processes $\gamma\gamma\rightarrow\pi^+\pi^-, K^+K^-$ in $k_{T}$ factorization would be a natural next step, and seems necessary because the higher-twist effects and the higher-power effects from the transverse momentum become important simultaneously in intermediate energy region. Compared with the previous calculations in collinear factorization~\cite{Gorsky:1989ev,Wang:2015mod}, the higher-twist calculations in $k_{T}$ factorization have several major improvements: Firstly, both the transverse momentum dependence of hadronic wave functions and that of hard kernels are included. The transverse momentum dependence of the hard kernel regularizes the end-point singularities of the internal propagators. While the transverse momentum dependence of the wave function makes a sizable suppression to the perturbative QCD contributions, and cannot be ignored as pointed out in Ref.~\cite{Jakob:1993iw}, especially in the few GeV region. Secondly, the corrections from the Sudakov resummation~\cite{Li:1992nu} substantially suppress the nonperturbative contributions from both the soft end-point regions and the large-$b$ regions. Moreover, the corrections from the threshold resummation~\cite{Kurimoto:2001zj} suppress the nonperturbative contributions from the end-point regions further and eliminate the end-point singularity without a artificial cut-off~\cite{Gorsky:1989ev} or the BHL prescription~\cite{Wang:2015mod} in the twist-3 calculations. Thirdly, the scales of the coupling constant $\alpha_s$ are chosen to be momentum-fraction dependent in order to avoid large logarithms from higher-order perturbative QCD corrections. All of the aforementioned improvements make the perturbative calculation become more self-consistent, even for momentum transfer as low as a few GeV.

In this paper, we present a detailed twist-3 calculations for the two-photon processes $\gamma\gamma\rightarrow\pi^+\pi^-, K^+K^-$ in the perturbative QCD approach based on the $k_{T}$ factorization theorem. At the twist-2 level, our results are consistent with the predictions given in Refs.~\cite{Vogt:1999sw,Vogt:2000bz}. While, with the twist-3 corrections, it is found that both the $\pi^{+}\pi^{-}$ and the $K^{+}K^{-}$ cross sections agree well with the experimental data~\cite{Heister:2003ae,Abe:2003vn,Nakazawa:2004gu}. This paper is organized as follows: In Sect.~\RNum{2}, the related conventions to these processes are introduced at first, next the twist-2 and twist-3 light-cone wave functions are described briefly, then we perform the calculation of the hard kernels with twist-3 corrections. The numerical analysis and discussion are given in Sect.~\RNum{3}. The last section is our summary and conclusion. The expression of the Sudakov function can be found in Appendix.

\section{Formalism}

\subsection{Kinematics and conventions}
\begin{figure}[!!htb]
\centering
\includegraphics[width=0.75\textwidth]{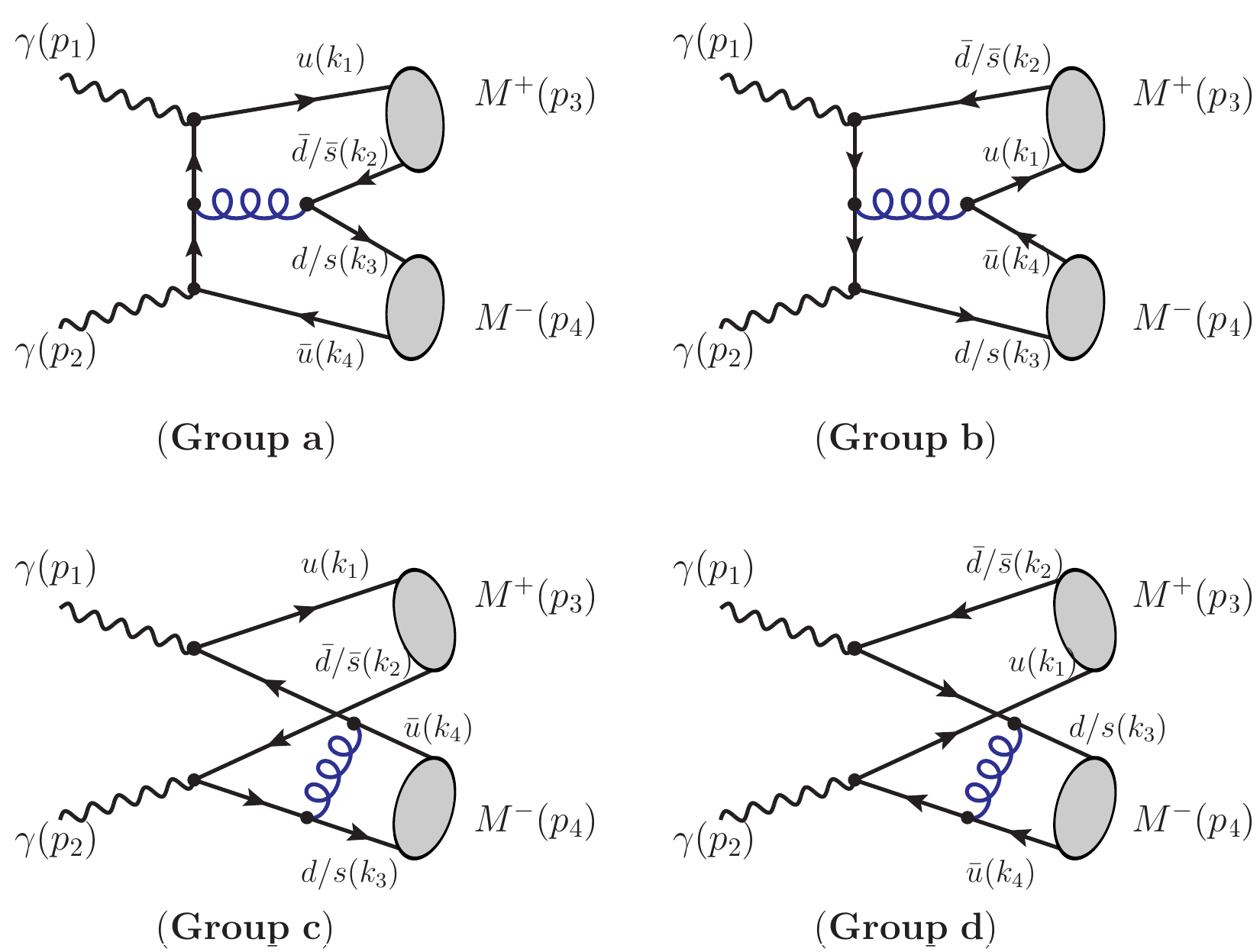}
\caption{Four basic Feynman diagrams for $\gamma\gamma\rightarrow M^{+}M^{-}(M=\pi,K)$. The total number of tree-level diagrams in groups $a$, $b$, $c$, and $d$ are 6, 6, 4, and 4, respectively. The grouping is based on the virtualities of the internal gluon.}
\label{feyndiag}
\end{figure}

We consider the simplest hadronic processes, $\gamma_1(p_1,\varepsilon_1^{\lambda_1})\gamma_2(p_2,\varepsilon_2^{\lambda_2})\rightarrow M^+(p_3)M^-(p_4)(M=\pi,K)$, where the incoming photons are real with momenta $p_1$, $p_2$ and polarization vectors $\varepsilon_1^{\lambda_1}$, $\varepsilon_2^{\lambda_2}$ with the photon  helicities $\lambda_1,\lambda_2=\pm1$, $M$ denotes the out-going pseudoscalar mesons with momenta $p_3$, $p_4$. The contributions to the $\gamma\gamma\rightarrow(q\bar{q})+(q\bar{q})$ amplitude arise from 20 tree-level Feynman diagrams, which can be grouped together into four basic types ($a$, $b$, $c$ and $d$) based on the virtualities of the internal gluon. We depict the four basic Feynman diagrams in Fig.~\ref{feyndiag}. The remaining diagrams arise from permutations of the gluon and quark lines.

For convenience, the amplitude is calculated in the center-of-momentum frame with the momenta of the final mesons along $z$-axis. $\theta$ is the scattering angle and $Q$ denotes the two-photon center-of-mass energy. In the light-cone coordinate, a momentum $A^\mu$ has the form $A^\mu=(A^+,A^-,\mathbf{A}_\perp)$ with $A^\pm=A^0\pm A^3$ and $\mathbf{A}_\perp=(A^1,A^2)$, the scalar product is defined as $A\cdot B=\frac{1}{2}(A^+B^-+A^-B^+)-\mathbf{A}_\perp\mathbf{B}_\perp$. Under these conventions, the momenta of photons and mesons read
\begin{align}
p_1=(2\omega c^2,2\omega s^2,\mathbf{p}),&\quad
p_2=(2\omega s^2,2\omega c^2,-\mathbf{p}),\nonumber\\
p_3=(2\omega,0,\mathbf{0}_\perp),&\quad
p_4=(0,2\omega,\mathbf{0}_\perp),
\end{align}
and the corresponding polarization vectors of the initial photons can be written as
\begin{align}
\varepsilon_1^+=\frac{1}{\sqrt{2}}(2sc,-2sc,s^2-c^2,-i),\quad
\varepsilon_1^-=\frac{1}{\sqrt{2}}(-2sc,2sc,c^2-s^2,-i),\nonumber\\
\varepsilon_2^+=\frac{1}{\sqrt{2}}(-2sc,2sc,c^2-s^2,-i),\quad
\varepsilon_2^-=\frac{1}{\sqrt{2}}(2sc,-2sc,s^2-c^2,-i)
\end{align}
with $\omega=\frac{Q}{2}$, $s=\sin\frac{\theta}{2}$, $c=\cos\frac{\theta}{2}$ and $\mathbf{p}=(2\omega sc,0)$ for abbreviation. The momenta of the quark and antiquark $k_1$, $k_2$ in the $M^+$ meson and $k_3$, $k_4$ in the $M^-$ meson as labeled in Fig.~\ref{feyndiag} can be written as
\begin{align}
&k_1=(2x\omega,0,\mathbf{k}_{\perp1}),\quad
k_2=(2\bar{x}\omega,0,-\mathbf{k}_{\perp1}),\nonumber\\
&k_3=(2y\omega,0,\mathbf{k}_{\perp2}),\quad
k_4=(2\bar{y}\omega,0,-\mathbf{k}_{\perp2})
\end{align}
with $\bar{x}\equiv1-x$ ($\bar{y}\equiv1-y$). Here $x$ ($y$) and $\bar{x}$ ($\bar{y}$) are the longitudinal momentum fractions of quark and antiquark, respectively, and the transverse momenta $\mathbf{k}_{\perp1}$, $\mathbf{k}_{\perp2}$ are relate to the momenta of the two out-going mesons.

\subsection{Light-cone wave functions}
The Fierz identity is generally employed to factorize the fermion flow:
\begin{align}\label{Fierz identity}
\overline{q}_{1\alpha}q_{2\beta}=&\frac{1}{4}I_{\beta\alpha}(\overline{q}_1q_2)
-\frac{1}{4}(i\gamma_5)_{\beta\alpha}(\overline{q}_1i\gamma^5q_2)
+\frac{1}{4}(\gamma_\mu)_{\beta\alpha}(\overline{q}_1\gamma^\mu q_2)\nonumber\\
&-\frac{1}{4}(\gamma_\mu\gamma_5)_{\beta\alpha}(\overline{q}_1\gamma^\mu\gamma^5q_2)
+\frac{1}{8}(\sigma_{\mu\nu}\gamma_5)_{\beta\alpha}(\overline{q}_1\sigma^{\mu\nu}\gamma^5q_2).
\end{align}
Here $q$ and $\bar{q}$ are the quark and antiquark field. The Dirac structure $\gamma^\mu\gamma^5$ in Eq.~(\ref{Fierz identity}) contributes at the twist-2 level, while $i\gamma^5$ and $\sigma^{\mu\nu}\gamma^5$ contribute at the twist-3 level.

With the light-cone expansion to twist-3 accuracy, the two-parton light-cone wave functions of $\pi^{-}$ can be expressed via the matrix element of the bilocal operator~\cite{Braun:1988qv,Braun:1989iv,Ball:1998je,Wei:2002iu}
\begin{align}
\label{matrix element}
\langle \pi^{-}(p)|\bar{d}_\beta(z_1)u_\alpha(z_2)|0\rangle=&\frac{if_\pi}{4}\int_0^1\mathrm{d}x\int \frac{\mathrm{d}^2\mathbf{k}_\perp}{16\pi^3} e^{i\left(xp\cdot z_1+\bar{x}p\cdot z_2-\mathbf{k}_\perp\cdot(\mathbf{z}_{1\perp}-\mathbf{z}_{2\perp})\right)}\nonumber\\
&\times\left\{\slashed{p}\gamma_5\Psi_\pi^\pi(x,\mathbf{k}_\perp)-\mu_{\pi}\gamma_5\left(\Psi_{\pi}^{p}(x,\mathbf{k}_\perp)
-\sigma_{\mu\nu}p^\mu (z_1-z_2)^\nu\frac{\Psi_\pi^\sigma(x,\mathbf{k}_\perp)}{6}\right)\right\}_{\alpha\beta},
\end{align}
where $f_\pi$ refers to the decay constant of pion and the mass parameter $\mu_\pi$ is defined as
\begin{equation}
\label{mupi}
\mu_\pi=\frac{m_\pi^2}{m_u+m_d},
\end{equation}
where $m_u$, $m_d$ are the current quark masses and $m_\pi$ denotes the pion meson mass. $\Psi_\pi^\pi$ is the twist-2 wave function and $\Psi_{\pi}^{p}$, $\Psi_{\pi}^{\sigma}$ are two-parton twist-3 wave functions. Due to the $SU(2)$ isotopic symmetry,
the wave functions of $\pi^{+}$ are identical with the wave functions of $\pi^{-}$. Generally, the twist-3 contributions are power-suppressed. However, compared with the twist-2 contributions, the suppression factor $\mu_\pi/Q$ in the two-parton twist-3 contributions is usually not small enough because of the chiral enhancement effects, i.e., through the relations between light quark masses in the chiral perturbation theory, $\mu_{\pi}(2~\mathrm{GeV})= 2.43~\mathrm{GeV}$~\cite{Khodjamirian:2009ys}. Especially,
in the intermediate energy region, such as the BELLE experimental range $2.4\ \mathrm{GeV}<Q<4.1\ \mathrm{GeV}$~\cite{Nakazawa:2004gu}, one can obtain the suppression factor $\mu_\pi/Q\sim \mathcal{O}(1)$. That is the reason why, in some processes~\cite{Gorsky:1989ev,Cao:1997st,Raha:2008ve,Raha:2010kz,Wang:2015mod,Cheng:2015qra,Cheng:2019ruz}, the two-parton twist-3 contributions could become even more important than the twist-2 contributions in a few $\mathrm{GeV}$ region.

Transforming the matrix element of the bilocal operator in Eq.~(\ref{matrix element}) from the coordinate space into the momentum space, one can obtain the light-cone projection operator~\cite{Wei:2002iu,Beneke:2000wa,Kroll:2018uvl}
\begin{equation}
M_{\alpha\beta}^\pi=\frac{if_\pi}{4}\left\{\slashed{p}\gamma_5\Psi_{\pi}^{\pi}-\mu_\pi\gamma_5\left(\Psi_{\pi}^{p}
-i\sigma_{\mu\nu}\frac{p^\mu \bar{p}^\nu}{p\cdot
\bar{p}}\frac{\Psi_{\pi}^{\sigma\prime}}{6}+i\sigma_{\mu\nu}p^\mu\frac{\Psi_{\pi}^{\sigma}}{6}\frac{\partial}{\partial k_{\bot\nu}}\right)\right\}_{\alpha\beta}
\end{equation}
with $\Psi_{\pi}^{\sigma\prime}=\partial\Psi_{\pi}^\sigma(x,\mathbf{k}_\perp)/\partial x$. From the first principles in QCD, the exact transverse momentum dependence of the above wave functions is still unknown. For the convenience of calculations, we assume that $\Psi_{\pi}^{\pi}$, $\Psi_{\pi}^{p}$ and $\Psi_{\pi}^{\sigma}$ have the same transverse momentum dependence. Following the ansatz made in Refs.~\cite{Jakob:1993iw,Bolz:1996wh}, the wave functions of the pion can be separated into two parts
\begin{align}
\Psi_{\pi}^i(x,\mathbf{k}_\perp,\mu)=\phi_{\pi}^i(x,\mu)\Sigma_{\pi}(x,\mathbf{k}_\perp),\quad (i=\pi,p,\sigma).
\end{align}
One part is the distribution amplitudes $\phi_{\pi}^{\pi}$, $\phi_{\pi}^{p}$ and $\phi_{\pi}^{\sigma}$. Using the conformal symmetry, the distribution amplitudes can be expanded in the Gegenbauer polynomials~\cite{Ball:1998je}
\begin{align}\label{pionDA}
\phi_{\pi}^{\pi}(x,\mu)=&6x\left(1-x\right)\left[1+a_2^\pi(\mu^2)C_2^{3/2}(2x-1)+a_4^\pi(\mu^2)C_4^{3/2}(2x-1)\right],\nonumber\\
\phi_{\pi}^{p}(x,\mu)=&1+\left(30\eta_{3\pi}(\mu^{2})-\frac{5}{2}\rho^{2}_{\pi}(\mu^{2})\right)C_{2}^{1/2}(2x-1)\nonumber\\
&\;\, +\left(-3\eta_{3\pi}(\mu^{2})\omega_{3\pi}(\mu^{2})-\frac{27}{20}\rho^{2}_{\pi}(\mu^{2})-\frac{81}{10}\rho^{2}_{\pi}(\mu^{2})
a_{2}^{\pi}(\mu^{2})\right)C_4^{1/2}(2x-1),\nonumber\\
\phi_{\pi}^{\sigma}(x,\mu)=&6x\left(1-x\right)\left[1+\left(5\eta_{3\pi}(\mu^{2})-\frac{1}{2}\eta_{3\pi}(\mu^{2})\omega_{3\pi}(\mu^{2})
-\frac{7}{20}\rho^{2}_{\pi}(\mu^{2})-\frac{3}{5}\rho^{2}_{\pi}(\mu^{2})
a_{2}^{\pi}(\mu^{2})\right)C_2^{3/2}(2x-1)\right]
\end{align}
with the normalization conditions
\begin{align}
\int_0^1\mathrm{d}x\phi_{\pi}^{\pi}(x,\mu)=1,\quad
\int_0^1\mathrm{d}x\phi_{\pi}^{p}(x,\mu)=1,\quad
\int_0^1\mathrm{d}x\phi_{\pi}^{\sigma}(x,\mu)=1.
\end{align}
Here $a_{2,4}^\pi$ are the Gegenbauer moments, the twist-3 parameters $\eta_{3\pi}$, $\rho_{\pi}^{2}$ are defined as
\begin{align}\label{pit3parameter}
\eta_{3\pi}=\frac{f_{3\pi}}{f_{\pi}\mu_{\pi}},\quad\quad
\rho_{\pi}^{2}=\frac{\left(m_{u}+m_{d}\right)^{2}}{m_{\pi}^{2}}
\end{align}
and the parameters $f_{3\pi}$, $\omega_{3\pi}$ are defined in terms of matrix elements of the local three-parton twist-3 operators~\cite{Ball:1998je}
\begin{align}
&\langle0|\bar{u}\sigma_{\mu\alpha}\gamma_{5}g_{s}G_{\nu\alpha}d|\pi^{-}(p)\rangle=2if_{3\pi} p_{\mu}p_{\nu},\nonumber\\
\langle0|\bar{u}\sigma_{\mu\alpha}\gamma_{5}\left[iD_{\beta},g_{s}G_{\nu\alpha}\right]&d-(3/7)i\partial_{\beta}\bar{u}\sigma_{\mu\alpha}\gamma_{5}g_{s} G_{\nu\alpha}d|\pi^{-}(p)\rangle=\frac{3}{14}if_{3\pi}\omega_{3\pi} p_{\mu}p_{\nu}p_{\beta}.
\end{align}
The other part is the $k_\perp$-dependence function $\Sigma_{\pi}(x,k_\perp)$.
With the constraint from the $\pi^{-}\rightarrow\mu^{-}\bar{\nu}_{\mu}$ decay~\cite{Lepage:1982gd}, one can obtain
\begin{align}
\int\frac{\mathrm{d}x\mathrm{d}^2\mathbf{k}_\perp}{16\pi^3}\Psi_{\pi}(x,\mathbf{k}_\perp,\mu)=1,
\end{align}
i.e., the function $\Sigma_{\pi}(x,\mathbf{k}_\perp)$ satisfy the normalization condition
\begin{align}
\int\frac{\mathrm{d}^2\mathbf{k}_\perp}{16\pi^3}\Sigma_{\pi}(x,\mathbf{k}_\perp)=1.
\end{align}
An appropriate candidate for the function $\Sigma_{\pi}(x,\mathbf{k}_\perp)$ is usually assumed to be a simple Gaussian form~\cite{Jakob:1993iw,Bolz:1997ez}
\begin{align}
\Sigma_{\pi}(x,\mathbf{k}_\perp)=\frac{16\pi^2\beta_{\pi}^2}{x(1-x)}\mathrm{exp}[-\frac{\beta_{\pi}^2 \mathbf{k}_\perp^2}{x(1-x)}],
\end{align}
where the oscillator parameter $\beta_{\pi}$ is determined by the mean square transverse momentum
\begin{align}
\langle \mathbf{k}_\perp^2\rangle_{\pi}=\frac{\int_0^1\mathrm{d}x\int \mathrm{d}^2\mathbf{k}_\perp \mathbf{k}_\perp^2|\Psi_{\pi}(x,\mathbf{k}_\perp)|^2}
{\int_0^1\mathrm{d}x\int \mathrm{d}^2\mathbf{k}_\perp|\Psi_{\pi}(x,\mathbf{k}_\perp)|^2}
=\frac{1}{2\beta_{\pi}^2}\frac{\int_0^1\mathrm{d}x|\phi_{\pi}(x)|^2}
{\int_0^1\mathrm{d}x\frac{|\phi_{\pi}(x)|^2}{x(1-x)}}.
\end{align}
The detailed discussions about the choice of the oscillator parameter $\beta_{\pi}$ are shown in Refs.~\cite{Jakob:1993iw,Bolz:1997ez}. In this work, we will take the value $\langle \mathbf{k}_\perp^2\rangle_{\pi}^{1/2}=0.35~\mathrm{GeV}$, which is compatible with the $\pi^{0}\rightarrow\gamma\gamma$ constraint~\cite{Lepage:1982gd}.

For the $K^{-}$ case, the matrix element of the bilocal operator can be expressed as~\cite{Ball:2006wn}
\begin{align}\label{hadronic matrix element}
\langle K^{-}(p)|\bar{s}_\beta(z_1)u_\alpha(z_2)|0\rangle=&\frac{if_K}{4}\int_0^1\mathrm{d}x\int\frac{\mathrm{d}^2\mathbf{k}_\perp}{16\pi^3} e^{i\left(xp\cdot z_1+\bar{x}p\cdot z_2-\mathbf{k}_\perp\cdot(\mathbf{z}_{1\perp}-\mathbf{z}_{2\perp})\right)}\nonumber\\
&\times\left\{\slashed{p}\gamma_5\Psi_{K}^{K}(x,\mathbf{k}_\perp)-\mu_{K}\gamma_5\left(\Psi_{K}^{p}(x,\mathbf{k}_\perp)
-\sigma_{\mu\nu}p^\mu(z_1-z_2)^\nu\frac{\Psi_{K}^{\sigma}(x,\mathbf{k}_\perp)}{6}\right)\right\}_{\alpha\beta},
\end{align}
where $f_{K}$ refers to the decay constant of kaon, and the chiral enhancement parameter is defined as $\mu_{K}=m_{K}^2/(m_u+m_s)$, $m_u$ and $m_s$ are the current quark masses and $m_K$ denotes the kaon meson mass.
We should note that the wave functions of $K^{+}$ are given by $\Psi_{K^{+}}^{i}(x,\mathbf{k}_\perp)
=\Psi_{K}^{i}(1-x,\mathbf{k}_\perp)$~\cite{Ball:2006wn}.

Similar to the pion case, the wave functions of the kaon meson can also be written as
\begin{align}
\Psi_{K}^{i}(x,\mathbf{k}_\perp,\mu)=\phi_{K}^{i}(x,\mu)\Sigma_{K}(x,\mathbf{k}_\perp),\quad (i=K,p,\sigma)
\end{align}
with
\begin{align}
\Sigma_{K}(x,\mathbf{k}_\perp)=\frac{16\pi^2\beta_{K}^2}{x(1-x)}\mathrm{exp}[-\frac{\beta_{K}^2 \mathbf{k}_\perp^2}{x(1-x)}].
\end{align}
Here we have adopted the approximation $\langle \mathbf{k}_\perp^2\rangle_{K}\approx\langle \mathbf{k}_\perp^2\rangle_{\pi}$.
The corresponding distribution amplitudes $\phi_{K}^{K}$, $\phi_{K}^{p}$ and $\phi_{K}^{\sigma}$ have the forms~\cite{Ball:2006wn}
\begin{align}\label{kaonDA}
\phi_{K}^{K}(x,\mu)&=
6x\left(1-x\right)\left[1+a_1^KC_1^{3/2}(2x-1)+a_2^KC_2^{3/2}(2x-1)\right],\nonumber\\
\phi_{K}^{p}(x,\mu)&=
1+3\rho_{+}^{K}\left(1+6a_{2}^{K}\right)-9\rho_{-}^{K}a_{1}^{K}
+\left[\frac{27}{2}\rho_{+}^{K}a_{1}^{K}-\rho_{-}^{K}\left(\frac{3}{2}+27a_{2}^{K}
\right)\right]C_{1}^{1/2}(2x-1)\nonumber\\
&+\left(30\eta_{3K}+15\rho^{K}_{+}a_{2}^{K}-3\rho_{-}^{K}a_{1}^{K}\right)
C_{2}^{1/2}(2x-1)
+\left(10\eta_{3K}\lambda_{3K}-\frac{9}{2}\rho^{K}_{-}a_{2}^{K}\right)C_{3}^{1/2}(2x-1)
-3\eta_{3K}\omega_{3K}C_{4}^{1/2}(2x-1)\nonumber\\
&+\frac{3}{2}\left(\rho^{K}_{+}
+\rho^{K}_{-}\right)\left(1-3a_{1}^{K}+6a_{2}^{K}\right)\ln x
+\frac{3}{2}\left(\rho^{K}_{+}-\rho^{K}_{-}\right)\left(1+3a_{1}^{K}
+6a_{2}^{K}\right)\ln(1-x),\nonumber\\
\phi_{K}^{\sigma}(x,\mu)&=
6x\left(1-x\right)\bigg{[}1+\frac{3}{2}\rho^{K}_{+}+15\rho^{K}_{+}a_{2}^{K}
-\frac{15}{2}\rho^{K}_{-}a_{1}^{K}
+\left(3\rho_{+}^{K}a_{1}^{K}-\frac{15}{2}\rho_{-}^{K}a_{2}^{K}
\right)C_{1}^{3/2}(2x-1)\nonumber\\
&+\left(5\eta_{3K}-\frac{1}{2}\eta_{3K}\omega_{3K}
+\frac{3}{2}\rho_{+}^{K}a_{2}^{K}\right)C_{2}^{3/2}(2x-1)
+\eta_{3K}\lambda_{3K}C_{3}^{3/2}(2x-1)\nonumber\\
&+\frac{3}{2}\left(\rho^{K}_{+}
+\rho^{K}_{-}\right)\left(1-3a_{1}^{K}+6a_{2}^{K}\right)\ln x
+\frac{3}{2}\left(\rho^{K}_{+}-\rho^{K}_{-}\right)\left(1+3a_{1}^{K}
+6a_{2}^{K}\right)\ln(1-x)
\bigg{]}
\end{align}
with the normalization conditions
\begin{align}
\int_0^1\mathrm{d}x\phi_{K}^{K}(x,\mu)=1,\quad
\int_0^1\mathrm{d}x\phi_{K}^{p}(x,\mu)=1,\quad
\int_0^1\mathrm{d}x\phi_{K}^{\sigma}(x,\mu)=1-\rho_{+}^{K}.
\end{align}
Here the $SU(3)$ symmetry breaking effects have already been taken into account. The parameters $a_{1,2}^{K}$ are the first two nonvanishing Gegenbauer moments, and the twist-3 parameters $\eta_{3K}$, $\rho_{\pm}^{K}$ read
\begin{align}\label{pit3parameter}
\eta_{3K}=\frac{f_{3K}}{f_{K}\mu_{K}},\quad\quad \rho_{+}^{K}=\frac{\left(m_{s}+m_{u}\right)^{2}}{m_{K}^{2}},
\quad\quad \rho_{-}^{K}=\frac{m_{s}^{2}-m_{u}^{2}}{m_{K}^{2}},
\end{align}
where the parameters $f_{3K}$, $\omega_{3K}$, $\lambda_{3K}$ are defined in terms of matrix elements of the local three-parton twist-3 operators~\cite{Ball:1998je}
\begin{flalign}
&\quad\quad\quad\quad\quad\quad\quad
\langle0|\bar{u}\sigma_{\mu\alpha}\gamma_{5}g_{s}G_{\nu\alpha}s|K^{-}(p)\rangle=2i f_{3K} p_{\mu}p_{\nu},\nonumber\\
&\langle0|\bar{u}\sigma_{\mu\alpha}\gamma_{5}\left[iD_{\beta},g_{s} G_{\nu\alpha}\right]s-(3/7)i\partial_{\beta} \bar{u} \sigma_{\mu\alpha}\gamma_{5} g_{s} G_{\nu\alpha}s|K^{-}(p)\rangle=\frac{3i}{14} f_{3K}\omega_{3K} p_{\mu}p_{\nu}p_{\beta},\nonumber\\
&\quad\quad\langle0|\bar{u}i\overleftarrow{D}_{\beta}\sigma_{\mu\alpha}\gamma_{5}g_{s}G_{\nu\alpha}s-\bar{u}\sigma_{\mu\alpha}\gamma_{5}g_{s} G_{\nu\alpha}i\overrightarrow{D}_{\beta} s|K^{-}(p)\rangle=\frac{i}{7}f_{3K}\lambda_{3K} p_{\mu}p_{\nu}p_{\beta}.
\end{flalign}

Using the leading-order renormalization group equations, the scale dependence of the various parameters is given by
\begin{align}
a_{n}^{\pi}(\mu^{2})&=L^{\gamma^{(0)}_{n}/\beta_{0}}a_{n}^{\pi}(\mu^{2}_{0}), &\mu_{\pi}(\mu^{2})&=L^{\gamma^{(0)}_{\mu_{\pi}}/\beta_{0}}\mu_{\pi}(\mu^{2}_{0}), &\rho_{\pi}^{2}(\mu^{2})&=L^{\gamma^{(0)}_{\rho_{\pi}^{2}}/\beta_{0}}\rho_{\pi}^{2}(\mu^{2}_{0}),\nonumber\\
f_{3\pi}(\mu^{2})&=L^{\gamma^{(0)}_{f_{3\pi}}/\beta_{0}}f_{3\pi}(\mu^{2}_{0}), &\eta_{3\pi}(\mu^{2})&=L^{\gamma^{(0)}_{\eta_{3\pi}}/\beta_{0}}\eta_{3\pi}(\mu^{2}_{0}), &\omega_{3\pi}(\mu^{2})&=L^{\gamma^{(0)}_{\omega_{3\pi}}/\beta_{0}}\omega_{3\pi}(\mu^{2}_{0}),
\end{align}
for the pion and
\begin{align}
a_{n}^{K}(\mu^{2})&=L^{\gamma^{(0)}_{n}/\beta_{0}}a_{n}^{K}(\mu^{2}_{0}),\quad \mu_{K}(\mu^{2})=L^{\gamma^{(0)}_{\mu_{K}}/\beta_{0}}\mu_{K}(\mu^{2}_{0}),\quad \rho_{+}^{K}(\mu^{2})=L^{\gamma^{(0)}_{\rho_{+}^{K}}/\beta_{0}}\rho_{+}^{K}(\mu^{2}_{0}),\quad \rho_{-}^{K}(\mu^{2})=L^{\gamma^{(0)}_{\rho_{-}^{K}}/\beta_{0}}\rho_{-}^{K}(\mu^{2}_{0}),\nonumber\\
f_{3K}(\mu^{2})&=L^{55/(36\beta_{0})}f_{3K}(\mu^{2}_{0})+\frac{2}{19}\left(L^{1/\beta_{0}}
-L^{55/(36\beta_{0})}\right)[f_{K}m_{s}](\mu^{2}_{0})+\frac{6}{65}\left(L^{55/(36\beta_{0})}
-L^{17/(9\beta_{0})}\right)[f_{K}m_{s}a_{1}^{K}](\mu^{2}_{0}),\nonumber\\
{[f_{3K}\omega_{3K}]}(\mu^{2})&=L^{26/(9\beta_{0})}[f_{3K}\omega_{3K}](\mu^{2}_{0})
+\frac{1}{170}\left(L^{1/\beta_{0}}
-L^{26/(9\beta_{0})}\right)[f_{K}m_{s}](\mu^{2}_{0})\nonumber\\
&+\frac{1}{10}\left(L^{17/(9\beta_{0})}
-L^{26/(9\beta_{0})}\right)[f_{K}m_{s}a_{1}^{K}](\mu^{2}_{0})+\frac{2}{15}\left(L^{43/(18\beta_{0})}
-L^{26/(9\beta_{0})}\right)[f_{K}m_{s}a_{2}^{K}](\mu^{2}_{0}),  \nonumber\\
{[f_{3K}\lambda_{3K}]}(\mu^{2})&=L^{37/(18\beta_{0})}[f_{3K}\lambda_{3K}](\mu^{2}_{0})
-\frac{14}{67}\left(L^{1/\beta_{0}}
-L^{37/(18\beta_{0})}\right)[f_{K}m_{s}](\mu^{2}_{0})\nonumber\\
&+\frac{14}{5}\left(L^{17/(9\beta_{0})}
-L^{37/(18\beta_{0})}\right)[f_{K}m_{s}a_{1}^{K}](\mu^{2}_{0})-\frac{4}{11}\left(L^{43/(18\beta_{0})}
-L^{37/(18\beta_{0})}\right)[f_{K}m_{s}a_{2}^{K}](\mu^{2}_{0}),
\end{align}
for the kaon with the evolution factor $L=\alpha_s(\mu^2)/\alpha_s(\mu_0^2)$ and $\beta_{0}=(33-2 n_{f})/12$.
To the leading logarithmic accuracy, the QCD running coupling constant can be expressed as
\begin{align}
\alpha_s(\mu^2)=\frac{\pi}{\beta_{0}\ln\left(\frac{\mu^2}{\Lambda_{QCD}^2}\right)},
\end{align}
and the anomalous dimensions are given by
\begin{align}
\gamma^{(0)}_{n}=&C_F\left(\psi(n+2)+\gamma_E-\frac{3}{4}-\frac{1}{2(n+1)(n+2)}\right),\nonumber\\
&\gamma^{(0)}_{\mu_{\pi}}=\gamma^{(0)}_{\mu_{K}}=-1,\quad
\gamma^{(0)}_{\rho_{\pi}^{2}}=\rho_{\pm}^{K}=2,\nonumber\\
\gamma^{(0)}_{f_{3\pi}}=\frac{7}{12}C_{F}+&\frac{1}{4}C_{A},\quad
\gamma^{(0)}_{\eta_{3\pi}}=\frac{4}{3}C_{F}+\frac{1}{4}C_{A},\quad
\gamma^{(0)}_{\omega_{3\pi}}=-\frac{7}{24}C_{F}+\frac{7}{12}C_{A},\nonumber\\
\end{align}
with $C_{F}=4/3$ and $C_{A}=3$.

In perturbative QCD approach, the convolution of wave functions and hard scattering kernels needs to be performed in the transverse configuration $b$-space. With the definition
\begin{align}
\hat{\Psi}(x,\mathbf{b},\mu)=\int\frac{\mathrm{d}^{2}\mathbf{k}_\perp}{(2\pi)^{2}}\Psi(x,\mathbf{k}_\perp,\mu)e^{i\mathbf{k}_\perp\cdot
\mathbf{b}},
\end{align}
one can obtain that the full soft wave functions of the pion and kaon in the transverse configuration $b$-space have the forms
\begin{align}
\hat{\Psi}_{M}^{i}(x,\mathbf{b},\mu)=\phi_{M}^{i}(x,\mu)\hat{\Sigma}_{M}(x,\mathbf{b}),\quad (M=\pi,K;\ i=\pi/K, p, \sigma)
\end{align}
where the Fourier transformed functions $\hat{\Sigma}_{M}(x,b)$ read
\begin{align}
\hat{\Sigma}_{M}(x,\mathbf{b})=\int\frac{\mathrm{d}^{2}\mathbf{k}_\perp}{(2\pi)^{2}}\Sigma_{M}(x,\mathbf{k}_\perp)e^{i\mathbf{k}_\perp\cdot
\mathbf{b}}=4\pi\ \mathrm{exp}[-\frac{x(1-x)\mathbf{b}^2}{4\beta_{M}^2}].
\end{align}

\subsection{The helicity amplitudes with twist-3 contributions }

In the perturbative QCD approach, the helicity amplitudes can be expressed as the convolution between the soft hadron wave functions and the hard scattering kernels with respect to both the longitudinal momentum fractions and the transverse separation of quark and antiquark. To the $\pi^+\pi^-$ production in two-photon collision, the helicity amplitudes have the form~\cite{Coriano:1998ge,Hsieh:2004ee}
\begin{align}\label{HA}
\mathcal{M}_{\lambda_1\lambda_2}{(Q,\theta)}=
\int_0^1\mathrm{d}x\mathrm{d}y\int\frac{\mathrm{d}^2\mathbf{b}_1}{4\pi}\frac{\mathrm{d}^2\mathbf{b}_{2}}{4\pi}
&\sum_{i,j=\pi,p,\sigma,\sigma^{\prime}}\hat{\Psi}_\pi^i(x,\mathbf{b}_1,\mu_F)
\hat{T}_{ij}^{\lambda_1\lambda_2}(x,y,Q,\theta,\mathbf{b}_{1},\mathbf{b}_{2},\mu_R)
\hat{\Psi}_\pi^j(y,\mathbf{b}_2,\mu_F)\nonumber\\
&\times S_t(x)S_t(y)\exp[-S(x,y,\mathbf{b}_1,\mathbf{b}_2,\mu_F,\mu_R)]
\end{align}
with
\begin{align}
\hat{\Psi}_\pi^{\sigma^{\prime}}(x,\mathbf{b}_1,\mu_F)
=\frac{\partial\hat{\Psi}_\pi^\sigma(x,\mathbf{b}_1,\mu_F)}{\partial x},\quad
\hat{\Psi}_\pi^{\sigma^{\prime}}(y,\mathbf{b}_2,\mu_F)
=\frac{\partial\hat{\Psi}_\pi^\sigma(y,\mathbf{b}_2,\mu_F)}{\partial y}.
\end{align}
$\hat{T}_{ij}^{\lambda_1\lambda_2}$ represent the Fourier transformed hard scattering kernels
\begin{align}\label{THHAT}
\hat{T}_{ij}^{\lambda_1\lambda_2}(x,y,Q,\theta,\mathbf{b}_{1},\mathbf{b}_{2},\mu_R)=&
\int\frac{\mathrm{d}^2\mathbf{k}_{\perp1}}{(2\pi)^2}\frac{\mathrm{d}^2\mathbf{k}_{\perp2}}{(2\pi)^2}
T_{ij}^{\lambda_1\lambda_2}(x,y,Q,\theta,\mathbf{k}_{\perp1},\mathbf{k}_{\perp2},\mu_R)\nonumber\\
&\times \mathrm{exp}[-i  \mathbf{k}_{\perp1}\cdot\mathbf{b}_1-i  \mathbf{k}_{\bot2}\cdot\mathbf{b}_2],
\end{align}
where $T_{ij}^{\lambda_1\lambda_2}$ with the transverse momentum $\mathbf{k}_{\perp1}$, $\mathbf{k}_{\perp2}$ kept can be calculated perturbatively by Feynman rules. In the helicity amplitudes, the scripts $i,j=\pi$ correspond to the twist-2$\times$twist-2 term, and the scripts $i,j=p,\sigma,\sigma^{\prime}$ correspond to the two-parton twist-3$\times$twist-3 terms. It is noteworthy that the two-parton twist-2$\times$twist-3 terms vanish because of the spin structure of the corresponding hard scattering kernels. Compared with the  twist-2$\times$twist-2 contributions, the contributions from the two-parton twist-3$\times$twist-3 terms are generally regarded as power-suppressed. However in a few GeV region,  the suppression factor is $\mu_{\pi}^{2}/Q^{2}\sim\mathcal{O}(1)$ as a consequence of the chiral enhancement effects. So the two-parton twist-3$\times$twist-3 contributions become important in this energy region.
While, compared with the twist-2$\times$twist-2 contributions, the contributions from the three-parton twist-3$\times$twist-3 terms and that from the two-parton twist-2$\times$twist-4 terms are suppressed by the factors $f_{3\pi}/(f_{\pi}Q)$ and $m_{\pi}^{2}/Q^{2}$
respectively. According to the values of the factors $f_{3\pi}/(f_{\pi}Q)$ and $m_{\pi}^{2}/Q^{2}$ in
the order $10^{-2}$ in the intermediate energy region, these two contributions are at least an order of magnitude smaller than the twist-2$\times$twist-2 contributions in this energy region. A similar situation can also be found in the calculations of the pion and kaon electromagnetic form factors~\cite{Raha:2010kz,Cheng:2019ruz}. In this work, we will neglect the contributions form the three-parton twist-3$\times$twist-3 terms and the two-parton twist-2$\times$twist-4 terms.

In the following calculation, we will show our procedure to the Fourier transformation of the hard scattering kernels by means of a specific Feynman diagram in the Group $\mathbf{a}$ in Fig.~\ref{feyndiag} as an example. The hard scattering kernels $T_{a_1ij}^{\lambda_1\lambda_2}$ with the subscript ``$a_1$" specifically refer to the results of one of the six diagrams in Group $\mathbf{a}$. Its leading-twist contributions are illustrated as
\begin{align}\label{TH}
T_{a_1\pi\pi}^{++}(x,y,Q,\theta,\mathbf{k}_{\perp1},\mathbf{k}_{\perp2},\mu_R)=&
\frac{1024}{81}\frac{i  \kappa_{1} s^2c^2\omega^4(xy+\bar{x}\bar{y})}
{(\tilde{q}_1^2+i \epsilon)(\tilde{q}_2^2+i \epsilon)(\tilde{g}^2+i \epsilon)},\nonumber\\
T_{a_1\pi\pi}^{+-}(x,y,Q,\theta,\mathbf{k}_{\perp1},\mathbf{k}_{\perp2},\mu_R)=&
\frac{1024}{81}\frac{i  \kappa_{1} s^2c^2\omega^4(\bar{x}y+x\bar{y})}
{(\tilde{q}_1^2+i \epsilon)(\tilde{q}_2^2+i \epsilon)(\tilde{g}^2+i \epsilon)}
\end{align}
with the factor $\kappa_{1}=\alpha\alpha_{s}(\mu_R)\pi^{2}f_{\pi}^{2}$. The corresponding quark and gluon propagators are
\begin{align}
\tilde{q}_1^2&=4s^2(c^2-\bar{x})\omega^2-(\mathbf{p}-\mathbf{k}_{\perp1})^2,\nonumber\\
\tilde{q}_2^2&=4s^2(c^2-y)\omega^2-(\mathbf{p}+\mathbf{k}_{\perp2})^2,\nonumber\\
\tilde{g}^2&=4\bar{x}y\omega^2-\mathbf{K}_{\perp}^2
\end{align}
with
\begin{equation}
\mathbf{K}_\perp=\mathbf{k}_{\perp1}-\mathbf{k}_{\perp2}.
\end{equation}
The convolution formula in Eq.~(\ref{HA}) involves two independent scales of this process, the factorization scale $\mu_F$ and the renormalization scale $\mu_R$. In analogy to the cases of the pion electromagnetic form factor~\cite{Li:1992nu,Raha:2008ve,Li:2010nn} and transition form factor~\cite{Kroll:1996jx,Li:2009pr,Kroll:2010bf}, the transverse separation $b_1$ and $b_2$ of the pion wave functions provide the factorization scales $\mu_{Fi}=1/b_i(i=1,2)$, below which the QCD dynamics is regarded as being nonperturbative and can be absorbed into the the soft wave function. While the renormalization scale $\mu_R=t$ is set to the largest mass scale associated with the momentum of the internal hard gluon propagators in order to minimize the higher-order QCD corrections.

To simplify the calculation, the hierarchy $Q^{2}\gg \bar{x} Q^{2}\sim y Q^{2}\gg \bar{x} y Q^{2},~\mathbf{k}_{1\perp}^{2},~\mathbf{k}_{2\perp}^{2}$ is postulated in the small-$\bar{x}$, $y$ region, as elaborated in Refs.~\cite{Li:2010nn,Hu:2012cp}.
Under this hierarchy, one can ignore the transverse momentum dependence of the internal quark propagators and retain it in the gluon propagator to regulate the endpoint singularity, then the hard kernels in Eq.~(\ref{TH}) are simplified as
\begin{align}\label{Thak}
T_{a_1\pi\pi}^{++}(x,y,Q,\theta,\mathbf{k}_{\perp1},\mathbf{k}_{\perp2},\mu_R)=
\frac{64}{81}\frac{i  \kappa_{1}c^2(xy+\bar{x}\bar{y})}
{s^2x\bar{y}}\frac{1}{(\tilde{g}^2+i \epsilon)},\nonumber\\
T_{a_1\pi\pi}^{+-}(x,y,Q,\theta,\mathbf{k}_{\perp1},\mathbf{k}_{\perp2},\mu_R)=
\frac{64}{81}\frac{i  \kappa_{1} c^2(\bar{x}y+x\bar{y})}
{s^2x\bar{y}}\frac{1}{(\tilde{g}^2+i \epsilon)},
\end{align}
where the terms contained the transverse momentum $\mathbf{k}_\perp$ in the numerators, which are power-suppressed compared with other $\mathcal{O}(Q^{2})$ terms~\cite{Kurimoto:2001zj}, have also been dropped. By using the Fourier transformation of the propagators regularized with the $i \epsilon$ prescription
\begin{equation}
\int\frac{\mathrm{d}^2\mathbf{k}_\perp}{(2\pi)^2}
\frac{\mathrm{exp}[-i \mathbf{k}_\perp\cdot\mathbf{b}]}{s-\mathbf{k}_\perp^2+i \epsilon}
=\left\{
\begin{array}{rcl}
-\frac{i }{4}\mathrm{H}_0^{(1)}(\sqrt{s}b) & &{\mbox{for}\quad s>0}\\
\\
-\frac{1}{2\pi}\mathrm{K}_0(\sqrt{-s}b)        & &{\mbox{for}\quad s<0}
\end{array}
\right. \, ,
\end{equation}
one can obtain the Fourier transformed hard scattering kernels~($b=|\mathbf{b}_{1}|$)
\begin{align}\label{THa}
\hat{T}_{a_1\pi\pi}^{++}(x,y,Q,\theta,\mathbf{b}_{1},\mathbf{b}_{2},\mu_R)=
\frac{16\kappa_{1}}{81}\frac{c^2(xy+\bar{x}\bar{y})}
{s^2x\bar{y}}\mathrm{H}_0^{(1)}(2\omega\sqrt{\bar{x}y}b)\delta^{2}(\mathbf{b}_{1}-\mathbf{b}_{2}),\nonumber\\
\hat{T}_{a_1\pi\pi}^{+-}(x,y,Q,\theta,\mathbf{b}_{1},\mathbf{b}_{2},\mu_R)=
\frac{16\kappa_{1}}{81}\frac{c^2(\bar{x}y+x\bar{y})}
{s^2x\bar{y}}\mathrm{H}_0^{(1)}(2\omega\sqrt{\bar{x}y}b)\delta^{2}(\mathbf{b}_{1}-\mathbf{b}_{2}).
\end{align}
Here $\mathrm{H}_0^{(1)}$ and $\mathrm{K}_0$ denote Hankel and modified Bessel function. As a consequence of the above simplification, the hard kernels in Eq.~(\ref{THa}) only depend on a single $b$ parameter. The underlying physics picture is that, the virtual quark lines involved in the hard kernels are thought of as being far from mass shell, and shrunk to a point~\cite{Li:1992nu,Coriano:1998ge}.

There are two types of resummation of the higher-order effects: the region with large transverse separation $b$ for Sudakov resummation and the small longitudinal momentum fraction $x$ region for threshold resummation. The Sudakov resummation of the double logarithms $\alpha_{s}\ln^{2}[x/(Q^{2}b^{2})]$ produced by overlapping
collinear and soft divergencies for massless quarks as well as the renormalization group equation transformation from the factorization scale $1/b$ to the renormalization scale $t$, are incorporated into the Sudakov factor $\mathrm{exp}[-S]$. In next-to-leading logarithm approximation, the Sudakov exponent $S$ reads~\cite{Li:1992nu,Dahm:1995ne,Coriano:1998ge}
\begin{align}\label{SE}
S(x,y,b,Q,t)=s(x,b,Q)+s(\bar{x},b,Q)+s(y,b,Q)+s(\bar{y},b,Q)
-\frac{2}{\beta_{0}}\mathrm{ln}\frac{\mathrm{ln}(t/\Lambda_{QCD})}{\mathrm{ln}(1/(b/\Lambda_{QCD}))},
\end{align}
where the function $s(x,b,Q)$ is originally derived by Botts and Sterman~\cite{Botts:1989kf} and later on slightly improved, for instance, by Li~\cite{Li:1992nu} and Kroll~\cite{Dahm:1995ne} et al. The explicit expression of $s(x,b,Q)$ is given in Appendix. One can find that the Sudakov factor exhibits a strong fall off at large $b$. And this property makes the nonperturbative contributions from large $b$, no matter what $x$ is, less important.

Since there exists the end-point enhancement in the two-parton twist-3 contributions (logarithmical enhancement) for the scattering process $\gamma\gamma\rightarrow\pi^{+}\pi^{-}$~\cite{Gorsky:1989ev,Wang:2015mod}, the Sudakov factor $\mathrm{exp}[-S]$ is still not effective enough to suppress the nonperturbative contributions from small $x$ region, which would spoil the perturbative calculation. It had been argued in Ref.~\cite{Kurimoto:2001zj} that as the end-point region is important, the corresponding large double logarithms $\alpha_{s}\mathrm{ln}^2x$ that arise from the higher-order corrections also need to be organized to all orders, and into a jet function $S_{t}(x)$ as a consequence of threshold resummation. In next-to-leading logarithm accuracy, the jet function $S_{t}(x)$ can be parameterized into a universal form~\cite{Kurimoto:2001zj,Li:2001ay}
\begin{equation}
S_t(x,Q)=\frac{2^{1+2c}\Gamma(3/2+c)}{\sqrt{\pi}\Gamma(1+c)}\left[x(1-x)\right]^{c}.
\end{equation}
In Ref.~\cite{Li:2009pr}, Li and Mishima have proposed a parabolic parametrization for the parameter $c$, and frozen $c$ at $c=1$, when it exceeds unity:
\begin{equation}
c=0.04Q^2-0.51Q+1.87.
\end{equation}
Since the factor $S_{t}(x)$ drops rapidly as $x\rightarrow0,1$, the end-point singularities are eliminated and the nonperturbative contributions from the dangerous end-point regions are suppressed effectively.

After including the above two types of resummation, both the twist-2 and twist-3 contributions are well-behaved, and that results in the perturbative calculation more self-consistent. Performing the integration over $\mathbf{b}_{2}$, the helicity amplitudes in Eq.~(\ref{HA}) can be then rewritten as a much clearer and simpler form in a single-$b$ form
\begin{align}\label{helicity amplitude}
\mathcal{M}_{\lambda_1\lambda_2}{(Q,\theta)}=&
\int_0^1\mathrm{d}x\mathrm{d}y\int\frac{\mathrm{d}^2\mathbf{b}}{(4\pi)^2}
\sum_{i,j=\pi,p,\sigma,\sigma^{\prime}}\hat{\Psi}_{\pi}^{i}(x,\mathbf{b},1/b)
\hat{T}_{\lambda_1\lambda_2}^{ij}(x,y,Q,\theta,\mathbf{b},t)
\hat{\Psi}_{\pi}^{j}(y,\mathbf{b},1/b)\nonumber\\
&\times S_t(x)S_t(y)\exp[-S(x,y,Q,b,t)].
\end{align}
Making the substitution $\int \mathrm{d}^2\mathbf{b}\rightarrow\int b\mathrm{d}b\mathrm{d}\varphi$ and performing the integral over the polar angle $\varphi$, we can obtain the helicity amplitudes for $\gamma\gamma\rightarrow\pi^+\pi^-$ process in the conjugate $b$ space by summing up all the 20 Feynman diagrams
\begin{align}\label{totla helicity amplitude}
\mathcal{M}_{\lambda_1\lambda_2}{(Q,\theta)}=&
\int_0^1\mathrm{d}x\mathrm{d}y\int\frac{b\mathrm{d}b}{(4\pi)^2}
\sum_{n=a}^{b,c,d}\Bigg(\sum_{i,j=\pi}^{p,\sigma,\sigma^{\prime}}\hat{\Psi}_{\pi}^{i}(x,b,1/b)
\hat{T}^{\lambda_1\lambda_2}_{nij}(x,y,Q,\theta,b,t_{n})
\hat{\Psi}_{\pi}^{j}(y,b,1/b)\nonumber\\
&\times S_t(x)S_t(y)\exp[-S(x,y,Q,b,t_{n})]\Bigg)
\end{align}
with the twist-2 hard kernels
\begin{align}
\hat{T}_{a\pi\pi}^{++}&=
\frac{16\kappa_{1}}{81}
\frac{xy+\bar{x}\bar{y}}{s^2c^2x\bar{y}}
\mathrm{F}(s_a,b), \nonumber \\
\hat{T}_{b\pi\pi}^{++}&=
\frac{4\kappa_{1}}{81}
\frac{xy+\bar{x}\bar{y}}{s^2c^2\bar{x}y}
\mathrm{F}(s_b,b), \nonumber \\
\hat{T}_{c\pi\pi}^{++}&=
-\frac{8\kappa_{1}}{81}
\left(\frac{xy+\bar{x}\bar{y}}{s^2c^2xy}+
\frac{(xy+\bar{x}\bar{y})(x-\bar{y})}{c^2x\bar{x}y\bar{y}}\right)
\mathrm{J}_0(-pb)\mathrm{F}(s_c,b), \nonumber \\
\hat{T}_{d\pi\pi}^{++}&=
-\frac{8\kappa_{1}}{81}
\left(\frac{xy+\bar{x}\bar{y}}{s^2c^2\bar{x}\bar{y}}+
\frac{(xy+\bar{x}\bar{y})(\bar{x}-y)}{c^2x\bar{x}y\bar{y}}\right)
\mathrm{J}_0(pb)\mathrm{F}(s_d,b),
\end{align}
\begin{align}
\hat{T}_{a\pi\pi}^{+-}&=
\frac{16\kappa_{1}}{81}
\left(4+\frac{s^{2}(\bar{x}y+x\bar{y})}{c^2x\bar{y}}
+\frac{c^{2}(\bar{x}y+x\bar{y})}{s^2x\bar{y}}\right)
\mathrm{F}(s_a,b), \nonumber \\
\hat{T}_{b\pi\pi}^{+-}&=
\frac{4\kappa_{1}}{81}
\left(4
+\frac{s^{2}(\bar{x}y+x\bar{y})}{c^2\bar{x}y}
+\frac{c^{2}(\bar{x}y+x\bar{y})}{s^2\bar{x}y}\right)
\mathrm{F}(s_b,b), \nonumber \\
\hat{T}_{c\pi\pi}^{+-}&=
\frac{8\kappa_{1}}{81}
\left(4
-\frac{s^2(\bar{x}y+x\bar{y})}{c^2\bar{x}\bar{y}}
-\frac{c^2(\bar{x}y+x\bar{y})}{s^2xy}
\right)
\mathrm{J}_0(-pb)\mathrm{F}(s_c,b), \nonumber \\
\hat{T}_{d\pi\pi}^{+-}&=
\frac{8\kappa_{1}}{81}
\left(4
-\frac{s^2(\bar{x}y+x\bar{y})}{c^2xy}
-\frac{c^2(\bar{x}y+x\bar{y})}{s^2\bar{x}\bar{y}}
\right)
\mathrm{J}_0(pb)\mathrm{F}(s_d,b),
\end{align}
and the nonzero twist-3 hard kernels
\begin{align}
\hat{T}_{app}^{++}&=
-\frac{8\kappa_{2}}{81}
\left(2
-\frac{1-\bar{x}y}{s^2x\bar{y}}
-\frac{1-\bar{x}y}{c^2x\bar{y}}\right)
\mathrm{F}(s_a,b),&\quad
\hat{T}_{app}^{+-}&=
\frac{8\kappa_{2}}{81}
\left(\frac{1}{s^2c^2}
-\frac{2(1-x\bar{y})}{\bar{x}y}\right)
\mathrm{F}(s_a,b),\nonumber \\
\hat{T}_{bpp}^{++}&=
-\frac{2\kappa_{2}}{81}
\left(2
-\frac{1-x\bar{y}}{s^2\bar{x}y}
-\frac{1-x\bar{y}}{c^2\bar{x}y}\right)
\mathrm{F}(s_b,b),&
\hat{T}_{bpp}^{+-}&=
\frac{2\kappa_{2}}{81}
\left(\frac{1}{s^2c^2}
-\frac{2(1-\bar{x}y)}{x\bar{y}}\right)
\mathrm{F}(s_b,b), \nonumber \\
\hat{T}_{cpp}^{++}&=
\frac{4\kappa_{2}}{81}
\left(2+\frac{1-\bar{x}\bar{y}}{s^2xy}
+\frac{1-xy}{c^2\bar{x}\bar{y}}\right)
\mathrm{J}_0(-pb)\mathrm{F}(s_c,b),&
\hat{T}_{cpp}^{+-}&=
\frac{4\kappa_{2}}{81}
\left(\frac{1}{s^2c^2}-2\right)
 \mathrm{J}_0(-pb)\mathrm{F}(s_c,b),\nonumber \\
\hat{T}_{dpp}^{++}&=
\frac{4\kappa_{2}}{81}
\left(2
+\frac{1-xy}{s^2\bar{x}\bar{y}}
+\frac{1-\bar{x}\bar{y}}{c^2xy}\right)
\mathrm{J}_0(pb)\mathrm{F}(s_d,b),&
\hat{T}_{dpp}^{+-}&=
\frac{4\kappa_{2}}{81}
\left(\frac{1}{s^2c^2}-2\right)
 \mathrm{J}_0(pb)\mathrm{F}(s_d,b),
\end{align}

\begin{align}
\hat{T}_{cp\sigma}^{++}&=
-\frac{2\kappa_{2}}{243}
\frac{ \omega b\left(s^2x+c^{2}\bar{x}\right)}{scx\bar{x}}
\mathrm{J}_{1}(-pb)\mathrm{F}(s_c,b),&\quad
\hat{T}_{cp\sigma}^{+-}&=
-\frac{2\kappa_{2}}{243}
\frac{ \omega b\left(s^2\bar{x}+c^{2}x+1\right)}{scx\bar{x}}
\mathrm{J}_{1}(-pb)\mathrm{F}(s_c,b), \nonumber \\
\hat{T}_{dp\sigma}^{++}&=
\frac{2\kappa_{2}}{243}
\frac{ \omega b\left(s^2\bar{x}+c^{2}x\right)}{scx\bar{x}}
\mathrm{J}_{1}(pb)\mathrm{F}(s_d,b),&
\hat{T}_{dp\sigma}^{+-}&=
\frac{2\kappa_{2}}{243}
\frac{\omega b\left(s^2x+c^{2}\bar{x}+1\right)}{scx\bar{x}}
\mathrm{J}_{1}(pb)\mathrm{F}(s_d,b),
\end{align}

\begin{align}
\hat{T}_{ap\sigma^{\prime}}^{++}&=
\frac{4\kappa_{2}}{243}
\left(\frac{1}{s^2c^2}
-\frac{2}{x}
\right)
\mathrm{F}(s_a,b), &\quad
\hat{T}_{ap\sigma^{\prime}}^{+-}&=
\frac{4\kappa_{2}}{243}
\left(\frac{\bar{x}^2}{s^2c^2x\bar{x}}
-\frac{2}{x\bar{x}}\right)
\mathrm{F}(s_a,b), \nonumber \\
\hat{T}_{bp\sigma^{\prime}}^{++}&=
-\frac{\kappa_{2}}{243}
\left(\frac{1}{s^2c^2}
-\frac{2}{\bar{x}}
\right)
\mathrm{F}(s_b,b),&
\hat{T}_{bp\sigma^{\prime}}^{+-}&=
-\frac{\kappa_{2}}{243}
\left(\frac{x^2}{s^2c^2x\bar{x}}
-\frac{2}{x\bar{x}}\right)
\mathrm{F}(s_b,b),  \nonumber \\
\hat{T}_{cp\sigma^{\prime}}^{++}&=
-\frac{2\kappa_{2}}{243}
\left(\frac{s^2-c^2}{s^2c^2}
-\frac{x-\bar{x}}{x\bar{x}}\right)
\mathrm{J}_0(-pb)\mathrm{F}(s_c,b), &
\hat{T}_{cp\sigma^{\prime}}^{+-}&=
-\frac{2\kappa_{2}}{243}
\left(\frac{s^2x^2-c^2\bar{x}^2}{s^2c^2x\bar{x}}
+\frac{x-\bar{x}}{x\bar{x}}\right)
\mathrm{J}_0(-pb)\mathrm{F}(s_c,b), \nonumber \\
\hat{T}_{dp\sigma^{\prime}}^{++}&=
-\frac{2\kappa_{2}}{243}
\left(\frac{c^2-s^2}{s^2c^2}
-\frac{x-\bar{x}}{x\bar{x}}\right)
\mathrm{J}_0(pb)\mathrm{F}(s_d,b),&
\hat{T}_{dp\sigma^{\prime}}^{+-}&=
-\frac{2\kappa_{2}}{243}
\left(\frac{c^2x^2-s^2\bar{x}^2}{s^2c^2x\bar{x}}
+\frac{x-\bar{x}}{x\bar{x}}\right)
\mathrm{J}_0(pb)\mathrm{F}(s_d,b),
\end{align}

\begin{align}
\hat{T}_{c\sigma p}^{++}&=
\frac{2\kappa_{2}}{243}
\frac{ \omega b\left(s^2y+c^{2}\bar{y}\right)}{scy\bar{y}}
\mathrm{J}_{1}(-pb)\mathrm{F}(s_c,b),&\quad
\hat{T}_{c\sigma p}^{+-}&=
\frac{2\kappa_{2}}{243}
\frac{ \omega b\left(s^2\bar{y}+c^{2}y+1\right)}{scy\bar{y}}
\mathrm{J}_{1}(-pb)\mathrm{F}(s_c,b),  \nonumber \\
\hat{T}_{d\sigma p}^{++}&=
-\frac{2\kappa_{2}}{243}
\frac{ \omega b\left(s^2\bar{y}+c^{2}y\right)}{scy\bar{y}}
\mathrm{J}_{1}(pb)\mathrm{F}(s_d,b),&
\hat{T}_{d\sigma p}^{+-}&=
-\frac{2\kappa_{2}}{243}
\frac{ \omega b\left(s^2y+c^{2}\bar{y}+1\right)}{scy\bar{y}}
\mathrm{J}_{1}(pb)\mathrm{F}(s_d,b),
\end{align}

\begin{align}
\hat{T}_{a\sigma^{\prime} p}^{++}&=
\frac{4\kappa_{2}}{243}
\left(\frac{1}{s^2c^2}-\frac{2}{\bar{y}}
\right)
\mathrm{F}(s_\mathrm{a},b), &\quad
\hat{T}_{a\sigma^{\prime} p}^{+-}&=
\frac{4\kappa_{2}}{243}
\left(\frac{y^2}{s^2c^2y\bar{y}}
-\frac{2}{y\bar{y}}\right)
\mathrm{F}(s_a,b),\nonumber \\
\hat{T}_{b\sigma^{\prime} p}^{++}&=
-\frac{\kappa_{2}}{243}
\left(\frac{1}{s^2c^2}-\frac{2}{y}
\right)
\mathrm{F}(s_\mathrm{b},b),&
\hat{T}_{b\sigma^{\prime} p}^{+-}&=
-\frac{\kappa_{2}}{243}
\left(\frac{\bar{y}^2}{s^2c^2y\bar{y}}
-\frac{2}{y\bar{y}}\right)
\mathrm{F}(s_b,b), \nonumber \\
\hat{T}_{c\sigma^{\prime} p}^{++}&=
\frac{2\kappa_{2}}{243}
\left(\frac{s^2-c^2}{s^2c^2}
-\frac{y-\bar{y}}{y\bar{y}}\right)
\mathrm{J}_0(-pb)\mathrm{F}(s_c,b),&
\hat{T}_{c\sigma^{\prime} p}^{+-}&=
\frac{2\kappa_{2}}{243}
\left(\frac{s^2y^2-c^2\bar{y}^2}{s^2c^2y\bar{y}}
+\frac{y-\bar{y}}{y\bar{y}}\right)
\mathrm{J}_0(-pb)\mathrm{F}(s_c,b),  \nonumber \\
\hat{T}_{d\sigma^{\prime} p}^{++}&=
\frac{2\kappa_{2}}{243}
\left(\frac{c^2-s^2}{s^2c^2}
-\frac{y-\bar{y}}{y\bar{y}}\right)
\mathrm{J}_0(pb)\mathrm{F}(s_d,b),&
\hat{T}_{d\sigma^{\prime} p}^{+-}&=
\frac{2\kappa_{2}}{243}
\left(\frac{c^2y^2-s^2\bar{y}^2}{s^2c^2y\bar{y}}
+\frac{y-\bar{y}}{y\bar{y}}\right)
\mathrm{J}_0(pb)\mathrm{F}(s_d,b),
\end{align}

\begin{align}
\hat{T}_{c\sigma\sigma}^{+-}&=
-\frac{2\kappa_{2}}{729}
\left(\frac{\omega^{2}b^2}{xy}+\frac{\omega^{2}b^2}{\bar{x}\bar{y}}\right)
\mathrm{J}_{2}(-pb)\mathrm{F}(s_c,b), \nonumber \\
\hat{T}_{d\sigma\sigma}^{+-}&=
-\frac{2\kappa_{2}}{729}
\left(\frac{\omega^{2}b^2}{xy}+\frac{\omega^{2} b^2}{\bar{x}\bar{y}}\right)
\mathrm{J}_{2}(pb)\mathrm{F}(s_d,b),
\end{align}

\begin{align}
\hat{T}_{c\sigma\sigma^{\prime}}^{++}&=
\frac{\kappa_{2}}{729}
\left(\frac{\omega b c}{sy}-\frac{\omega b s}{c\bar{y}}\right)
\mathrm{J}_{1}(-pb)\mathrm{F}(s_c,b),&\quad
\hat{T}_{c\sigma\sigma^{\prime}}^{+-}&=
\frac{\kappa_{2}}{729}
\left(\frac{\omega bc(1+\bar{x})}{sxy}-\frac{\omega bs(1+x)}{c\bar{x}\bar{y}}\right)
\mathrm{J}_{1}(-pb)\mathrm{F}(s_c,b),  \nonumber \\
\hat{T}_{d\sigma\sigma^{\prime}}^{++}&=
\frac{\kappa_{2}}{729}
\left(\frac{\omega b c}{s\bar{y}}-\frac{\omega b s}{cy}\right)
\mathrm{J}_{1}(pb)\mathrm{F}(s_d,b),&
\hat{T}_{d\sigma\sigma^{\prime}}^{+-}&=
\frac{\kappa_{2}}{729}
\left(\frac{\omega bc(1+x)}{s\bar{x}\bar{y}}-\frac{\omega bs(1+\bar{x})}{cxy}\right)
\mathrm{J}_{1}(pb)\mathrm{F}(s_d,b),
\end{align}

\begin{align}
\hat{T}_{c\sigma^{\prime}\sigma}^{++}&=
\frac{\kappa_{2}}{729}
\left(\frac{\omega b c}{sx}-\frac{\omega b s}{c\bar{x}}\right)
\mathrm{J}_{1}(-pb)\mathrm{F}(s_c,b),&\quad
\hat{T}_{c\sigma^{\prime}\sigma}^{+-}&=
\frac{\kappa_{2}}{729}
\left(\frac{\omega bc(1+\bar{y})}{sxy}-\frac{\omega bs(1+y)}{c\bar{x}\bar{y}}\right)
\mathrm{J}_{1}(-pb)\mathrm{F}(s_c,b), \nonumber \\
\hat{T}_{d\sigma^{\prime}\sigma}^{++}&=
\frac{\kappa_{2}}{729}
\left(\frac{\omega b c}{s\bar{x}}-\frac{\omega b s}{cx}\right)
\mathrm{J}_{1}(pb)\mathrm{F}(s_d,b),&
\hat{T}_{d\sigma^{\prime}\sigma}^{+-}&=
\frac{\kappa_{2}}{729}
\left(\frac{\omega bc(1+y)}{s\bar{x}\bar{y}}-\frac{\omega bs(1+\bar{y})}{cxy}\right)
\mathrm{J}_{1}(pb)\mathrm{F}(s_d,b),
\end{align}

\begin{align}
\hat{T}_{a\sigma^{\prime}\sigma^{\prime}}^{++}=&
\frac{2\kappa_{2}}{729}
\left(\frac{s^{2}(1-\bar{x}y)}{c^2x\bar{y}}
+\frac{c^{2}(1-\bar{x}y)}{s^2x\bar{y}}\right)
\mathrm{F}(s_a,b), \nonumber \\
\hat{T}_{b\sigma^{\prime}\sigma^{\prime}}^{++}=&
\frac{\kappa_{2}}{1458}
\left(\frac{s^{2}(1-x\bar{y})}{c^2\bar{x}y}
+\frac{c^{2}(1-x\bar{y})}{s^2\bar{x}y}\right)
\mathrm{F}(s_b,b), \nonumber \\
\hat{T}_{c\sigma^{\prime}\sigma^{\prime}}^{++}=&
-\frac{\kappa_{2}}{729}
\left(\frac{s^2(1-xy)}{c^2\bar{x}\bar{y}}
+\frac{c^2(1-\bar{x}\bar{y})}{s^2xy}\right)
 \mathrm{J}_0(-pb)\mathrm{F}(s_c,b), \nonumber \\
\hat{T}_{d\sigma^{\prime}\sigma^{\prime}}^{++}=&
-\frac{\kappa_{2}}{729}
\left(\frac{s^2(1-\bar{x}\bar{y})}{c^2xy}
+\frac{c^2(1-xy)}{s^2\bar{x}\bar{y}}\right)
\mathrm{J}_0(pb)\mathrm{F}(s_d,b),
\end{align}

\begin{align}
\hat{T}_{a\sigma^{\prime}\sigma^{\prime}}^{+-}=&
\frac{2\kappa_{2}}{729}
\left(\frac{s^{2}(1+\bar{x}y)}{c^2x\bar{y}}
+\frac{c^{2}(1+\bar{x}y)}{s^2x\bar{y}}\right)
\mathrm{F}(s_a,b), \nonumber \\
\hat{T}_{b\sigma^{\prime}\sigma^{\prime}}^{+-}=&
\frac{\kappa_{2}}{1458}
\left(\frac{s^{2}(1+x\bar{y})}{c^2\bar{x}y}
+\frac{c^2(1+x\bar{y})}{s^2\bar{x}y}\right)
\mathrm{F}(s_b,b), \nonumber \\
\hat{T}_{c\sigma^{\prime}\sigma^{\prime}}^{+-}=&
-\frac{\kappa_{2}}{729}
\left(\frac{s^2(1+xy)}{c^2\bar{x}\bar{y}}
+\frac{c^2(1+\bar{x}\bar{y})}{s^2xy}\right)
\mathrm{J}_0(-pb)\mathrm{F}(s_c,b), \nonumber \\
\hat{T}_{d\sigma^{\prime}\sigma^{\prime}}^{+-}=&
-\frac{\kappa_{2}}{729}
\left(\frac{s^2(1+\bar{x}\bar{y})}{c^2xy}
+\frac{c^2(1+xy)}{s^2\bar{x}\bar{y}}\right)
\mathrm{J}_0(pb)\mathrm{F}(s_d,b).
\end{align}
Here the factor $\kappa_{2}$ is defined as
\begin{align}
\kappa_{2}=\frac{\mu_{\pi}^2}{\omega^{2}}\kappa_{1}
=\frac{\mu_{\pi}^2}{\omega^{2}}\alpha\alpha_{s}(t_n)\pi^{2}f_{\pi}^{2}.
\end{align}
Compared with the twist-2 hard kernels, the twist-3 ones are easily found to be suppressed by the factor $\kappa_{2}/\kappa_{1}=\mu_{\pi}^2/\omega^{2}$.
In the above expressions for the $\hat{T}^{\lambda_1\lambda_2}_{nij}$, $\mathrm{J}_0$ represents the Bessel function with the variable $pb=2\omega scb$ and the function $\mathrm{F}(s_i,b)$ is defined as
\begin{equation}
\mathrm{F}(s_i,b)
=\left\{
\begin{array}{rcl}
2\pi \mathrm{H}_0^{(1)}(\sqrt{s_i}b) & &{\mbox{for}\quad s_i>0}\\
\\
-4i \mathrm{K}_0(\sqrt{-s_i}b)        & &{\mbox{for}\quad s_i<0}
\end{array}
\right. \,.
\end{equation}
The renormalization scales $t_n$ ($n=a,b,c,d$), entering $\hat{T}_H$ as the arguments of strong coupling constant $\alpha_s$ and depending on the kinematics of the specific group, are chosen as
\begin{align}
t_a=\mathrm{max}\left\{2\omega\sqrt{\bar{x}y},1/b\right\},& \quad
t_b=\mathrm{max}\left\{2\omega\sqrt{x\bar{y}},1/b\right\}, \nonumber \\
t_c=\mathrm{max}\left\{2\omega\sqrt{(s^2(\bar{x}-y)-\bar{x}\bar{y})},1/b\right\},& \quad
t_d=\mathrm{max}\left\{2\omega\sqrt{(s^2(x-\bar{y})-xy)},1/b\right\}.
\end{align}
The factors $s_n$ ($n=a,b,c,d$) come from the gluon propagators in each group
\begin{align}
s_a=4\bar{x}y\omega^2,& \quad
s_b=4x\bar{y}\omega^2, \nonumber \\
s_c=4(\bar{x}-s^2)(y-c^2)\omega^2,& \quad
s_d=4(x-s^2)(\bar{y}-c^2)\omega^2.
\end{align}
In addition, it is worth noticing the hard kernels with different helicity have the relations
\begin{align}
\hat{T}_{nij}^{++}=\hat{T}_{nij}^{--},\quad \hat{T}_{nij}^{+-}=\hat{T}_{nij}^{-+}.
\end{align}
For the $\gamma\gamma\rightarrow K^+K^-$ process, the hard kernels can be obtained directly with the replacements of $e_d\rightarrow e_s$, $f_\pi\rightarrow f_K$, $\mu_\pi\rightarrow\mu_K$. And the corresponding helicity amplitudes can be expressed as
\begin{align}\label{Kaon helicity amplitude}
\mathcal{M}^{\prime}_{\lambda_1\lambda_2}{(Q,\theta)}=&
\int_0^1\mathrm{d}x\mathrm{d}y\int\frac{b\mathrm{d}b}{(4\pi)^2}
\sum_{n=a}^{b,c,d}\Bigg(\sum_{i,j=K}^{p,\sigma,\sigma^{\prime}}\hat{\Psi}_{K}^{i}(1-x,b,1/b)
\hat{T}^{\lambda_1\lambda_2}_{nij}(x,y,Q,\theta,b,t_{n})
\hat{\Psi}_{K}^{j}(y,b,1/b)\nonumber\\
&\times S_t(x)S_t(y)\exp[-S(x,y,Q,b,t_{n})]\Bigg).
\end{align}
with
\begin{align}
\hat{\Psi}_{K}^{\sigma^{\prime}}(1-x,\mathbf{b}_1,\mu_F)
=\frac{\partial\hat{\Psi}_{K}^\sigma(1-x,\mathbf{b}_1,\mu_F)}{\partial x},\quad
\hat{\Psi}_{K}^{\sigma^{\prime}}(y,\mathbf{b}_2,\mu_F)
=\frac{\partial\hat{\Psi}_{K}^\sigma(y,\mathbf{b}_2,\mu_F)}{\partial y}.
\end{align}
\section{Numerical analysis}

The differential cross sections of the processes $\gamma\gamma\rightarrow M^+M^-$ $(M=\pi,K)$ can be expressed as
\begin{equation}
\frac{\mathrm{d}\sigma(\gamma\gamma\rightarrow M^+ M^-)}{\mathrm{d}|\cos\theta|}=
\frac{1}{32\pi Q^2}\frac{1}{4}\sum_{\lambda_1,\lambda_2=\pm1}\mid\mathcal{M}_{\lambda_1\lambda_2}\mid^2.
\end{equation}
In order to make comparison with the experimental measurements, we integrate over the scattering angle in the region $|\cos\theta|<0.6$ to get the cross sections $\sigma_0$ for both $\pi^+\pi^-$ and $K^+K^-$ processes.
In this work, we focus on the intermediate energy region $1.0~\mathrm{GeV}<Q<7.0~\mathrm{GeV}$ to check the applicability of the perturbative QCD in $\gamma\gamma\rightarrow\pi^{+}\pi^{-},~K^{+}K^{-}$ processes.

\begin{table}[!!htb]
\caption{The nonperturbative input parameters of the pion and kaon distribution amplitudes in our calculations.}
\label{tab:parameters}
\begin{center}
\scalebox{1.05}{\begin{tabular}{|c|c||c|c||c|}
\hline
$\pi$              &$\mu=1\,\mathrm{GeV}$   &$K$              &$\mu=1\,\mathrm{GeV}$   &$\mathrm{units/Refs.}$                  \nonumber\\
\toprule
$a_2^\pi$          &$0.17\pm0.08$           &$a_1^K$          &$0.10\pm0.04$           &\cite{Khodjamirian:2009ys,Khodjamirian:2011ub}                      \nonumber\\
$a_4^\pi$          &$0.06\pm0.10$           &$a_2^K$          &$0.25\pm0.15$           &\cite{Khodjamirian:2009ys,Khodjamirian:2011ub}                         \nonumber\\
\hline
\hline
$\pi$              &$\mu=2\,\mathrm{GeV}$   &$K$              &$\mu=2\,\mathrm{GeV}$   &$\mathrm{units/Refs.}$                  \nonumber\\
\toprule
$\mu_\pi$          &$2.50\pm0.30$           &$\mu_K$          &$2.49\pm0.26$           &$\mathrm{GeV}$,   \cite{Khodjamirian:2017fxg}   \nonumber\\
$f_{3\pi}$         &$0.0031$                &$f_{3K}$         &$0.0033$                &$\mathrm{GeV}^2$, \cite{Ball:2006wn} \nonumber\\
$\omega_{3\pi}$    &$-1.1$                  &$\omega_{3K}$    &$-0.9$                  &\cite{Ball:2006wn}                           \nonumber\\
$\lambda_{3\pi}$   &$0$                     &$\lambda_{3K}$   &$1.45$                  &\cite{Ball:2006wn}                           \nonumber\\
\hline
\end{tabular}}
\end{center}
\end{table}

In one-loop accuracy, the QCD running coupling is given by $\alpha_s(\mu^2)=\pi/[\beta_{0}\ln(\mu^2/\Lambda_{QCD}^2)]$ with the QCD scale $\Lambda_{QCD}=0.2\ \mathrm{GeV}$ and the interaction scale $\mu=t_n$ ($n=a,b,c,d$) corresponding to each group in Fig.~\ref{feyndiag}.
The most important parameters used in our numerical analysis are listed in Table.~\ref{tab:parameters}. In addition, the decay constants are taken as $f_{\pi}=130.4~\mathrm{MeV}$ and $f_{K}=159.8~\mathrm{MeV}$. For the meson masses and the quark mass entering the pion and kaon distribution amplitudes, we take $m_\pi =139.6~\mathrm{MeV}$, $m_K=493.7~\mathrm{MeV}$ and $m_{s}=95~\mathrm{MeV}$ in $\overline{\mathrm{MS}}$ scheme. All of the values are quoted from PDG~\cite{Tanabashi:2018oca}. The Gegenbauer moments $a_{2,4}^\pi$ and $a_{1,2}^K$ in the pion and kaon distribution amplitudes come from QCD sum rule~\cite{Khodjamirian:2009ys,Khodjamirian:2011ub}. For the chiral enhancement parameters $\mu_\pi$ and $\mu_K$, we employ the well known chiral perturbative theory relations~\cite{Leutwyler:1996qg}
\begin{align}
\mathcal{R}=\frac{2m_{s}}{m_{u}+m_{d}}=24.4\pm1.5,\quad\quad
\mathcal{Q}^2=\frac{m_{s}^{2}-(m_{u}+m_{d})^{2}/4}{m_{d}^{2}-m_{u}^{2}}=(22.7\pm0.8)^{2}
\end{align}
and obtain
\begin{align}
\mu_\pi=\frac{m_\pi^2\mathcal{R}}{2m_s},\quad\quad
\mu_K=\frac{m_K^2}{m_s\left[1+\frac{1}{\mathcal{R}}\left(1-\frac{\mathcal{R}^2-1}{4\mathcal{Q}^2}\right)\right]}.
\end{align}
The remaining parameters of the twist-3 distribution amplitudes in our calculations are taken from Ref.~\cite{Ball:2006wn} and shown in Table.~\ref{tab:parameters} for brevity.

\begin{table}[!!htbp]
\caption{Gegenbauer moments of five sample models of the pion and kaon distribution amplitudes at the scale $\mu_0=1\ \mathrm{GeV}$.}
\label{tab:Geg-moments}
\scalebox{1.05}{\begin{tabular}{cccccccc}
\hline\hline
  &\multicolumn{3}{c}{$\quad$$\gamma\gamma\rightarrow\pi^+\pi^-$} &$\quad$  &\multicolumn{3}{c}{$\gamma\gamma\rightarrow K^+K^-$} \\
                      \cline{3-4}                                  \cline{6-8}
Model           &$\;$  &$a_2^\pi$   &$a_4^\pi$   &$\quad$   &$a_1^K$   &$\,$   &$a_2^K$  \\
\hline
\RNum{1}  &$\;$  &$0.17$      &$0.06$      &$\quad$   & $0.10$   &$\,$   &$0.25$   \\
\RNum{2}  &$\;$  &$0.09$      &$-0.04$     &$\quad$   & $0.06$   &$\,$   &$0.10$   \\
\RNum{3}  &$\;$  &$0.25$      &$0.16$      &$\quad$   & $0.14$   &$\,$   &$0.40$   \\
\RNum{4}  &$\;$  &$0.09$      &$0.16$      &$\quad$   & $0.06$   &$\,$   &$0.40$   \\
\RNum{5}  &$\;$  &$0.25$      &$-0.04$     &$\quad$   & $0.14$   &$\,$   &$0.10$   \\
\hline
\hline
\end{tabular}}
\end{table}

\begin{figure}[!!htb]
\centering
\includegraphics[width=0.47\textwidth]{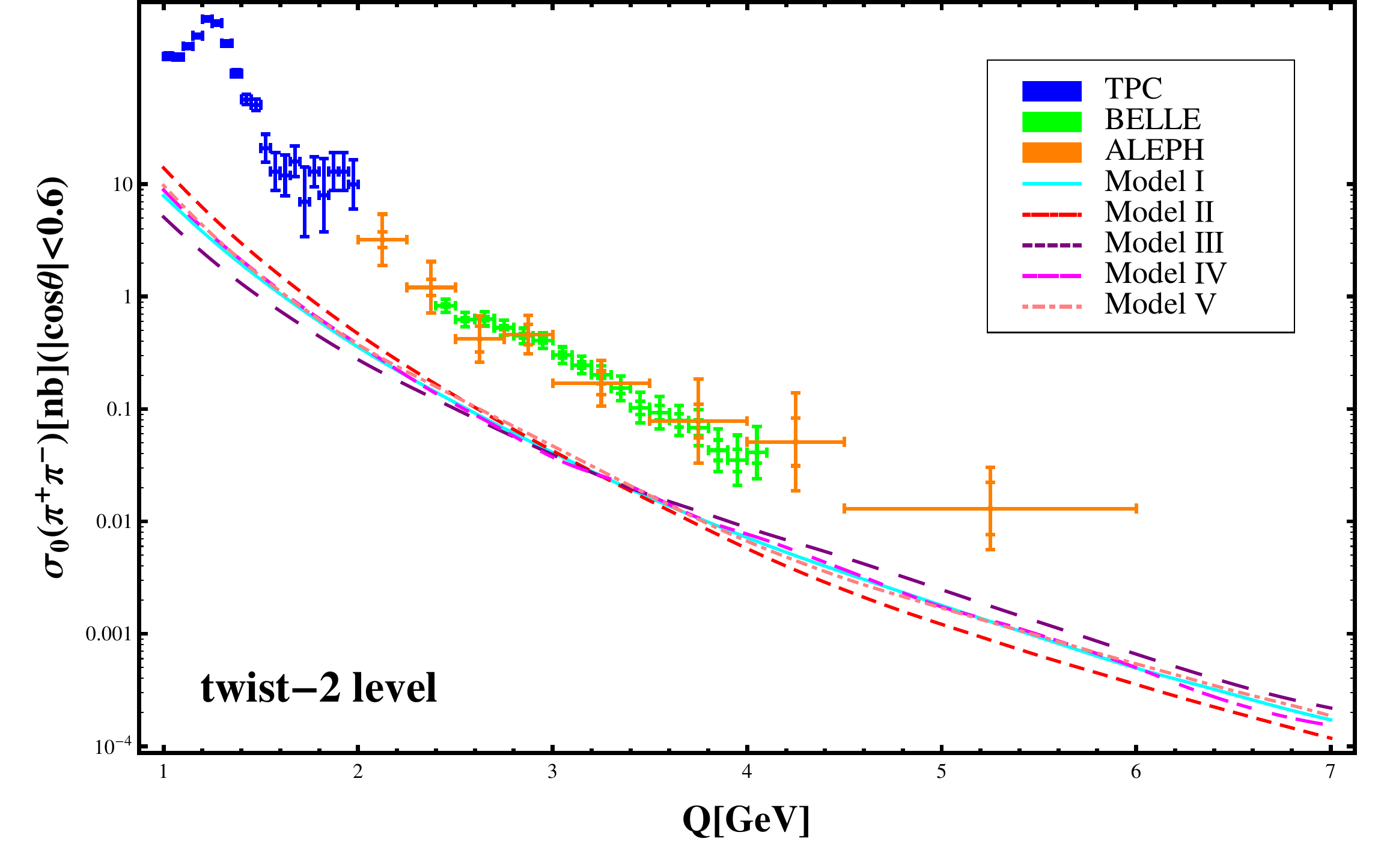}\hspace{0.5cm}
\includegraphics[width=0.47\textwidth]{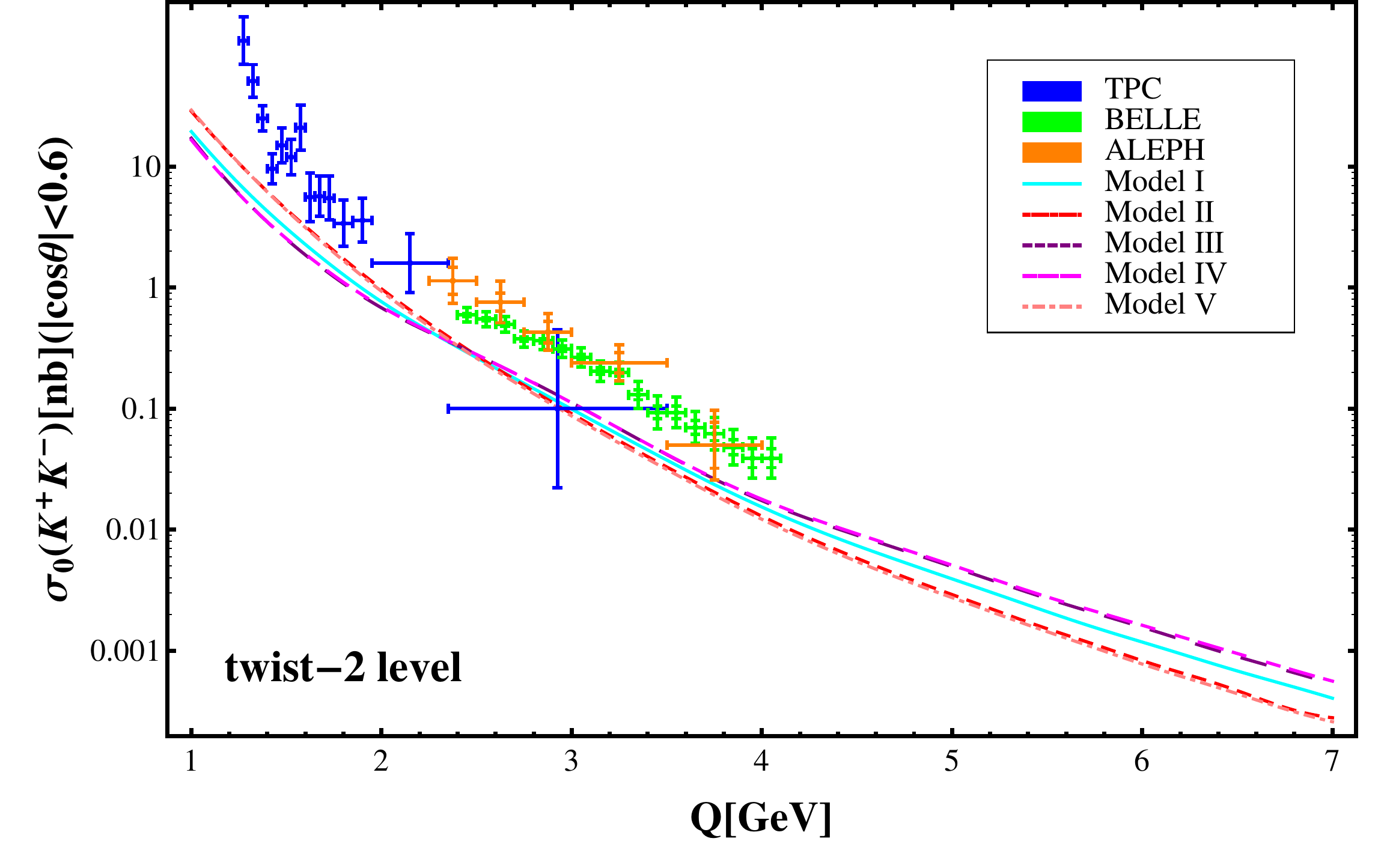}
\caption{Twist-2 results of the cross sections $\sigma_{0}(\pi^{+}\pi^{-})$ and $\sigma_{0}(K^{+}K^{-})$ with the five sample models of the distribution amplitudes listed in Table.~\ref{tab:Geg-moments}. The points with errors are the experimental data from Refs.~\cite{Aihara:1986qk,Heister:2003ae,Nakazawa:2004gu}.}
\label{fig:fig2}
\end{figure}

\begin{figure}[!!htb]
\centering
\includegraphics[width=0.32\textwidth]{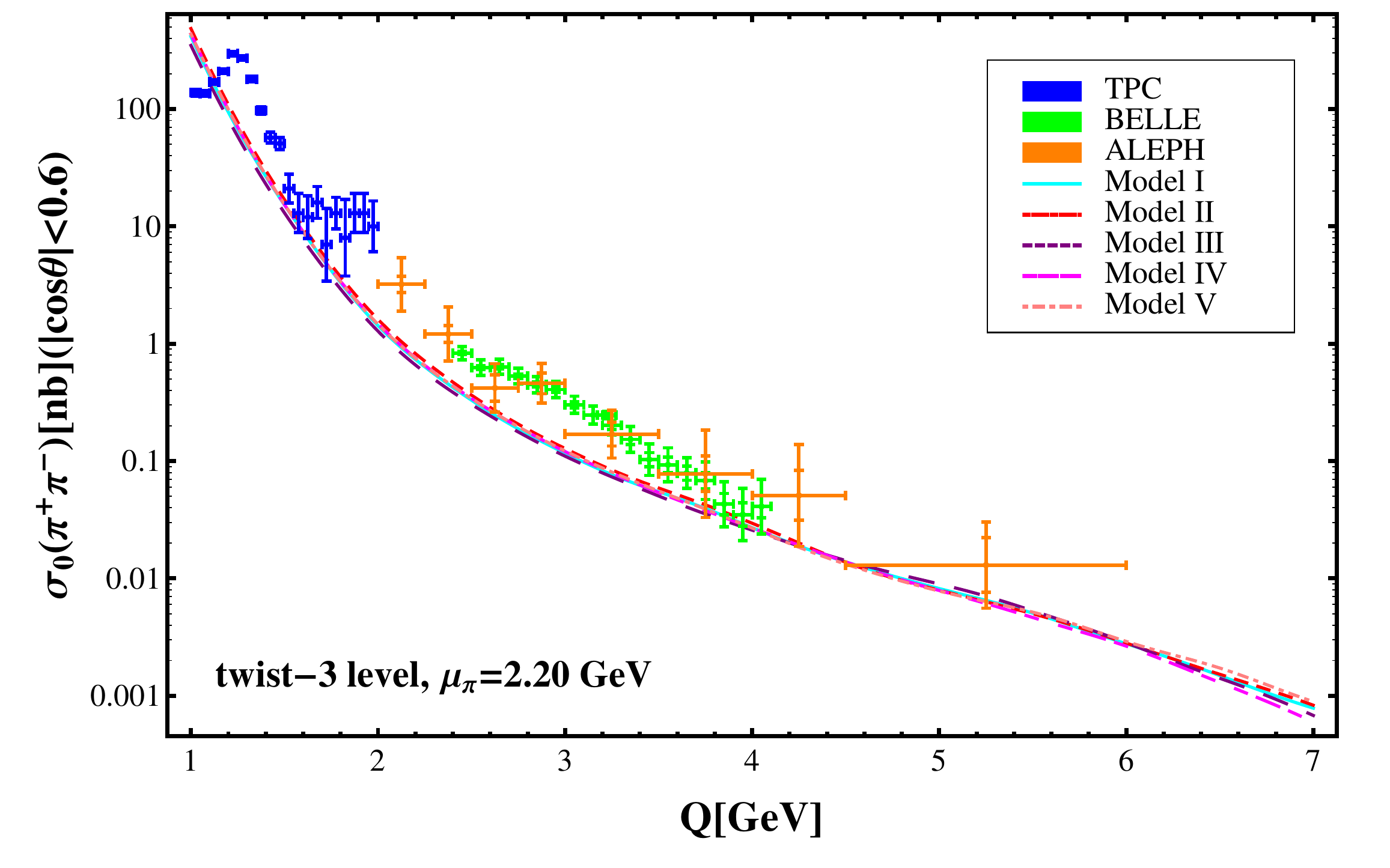}\hspace{0.1cm}
\includegraphics[width=0.32\textwidth]{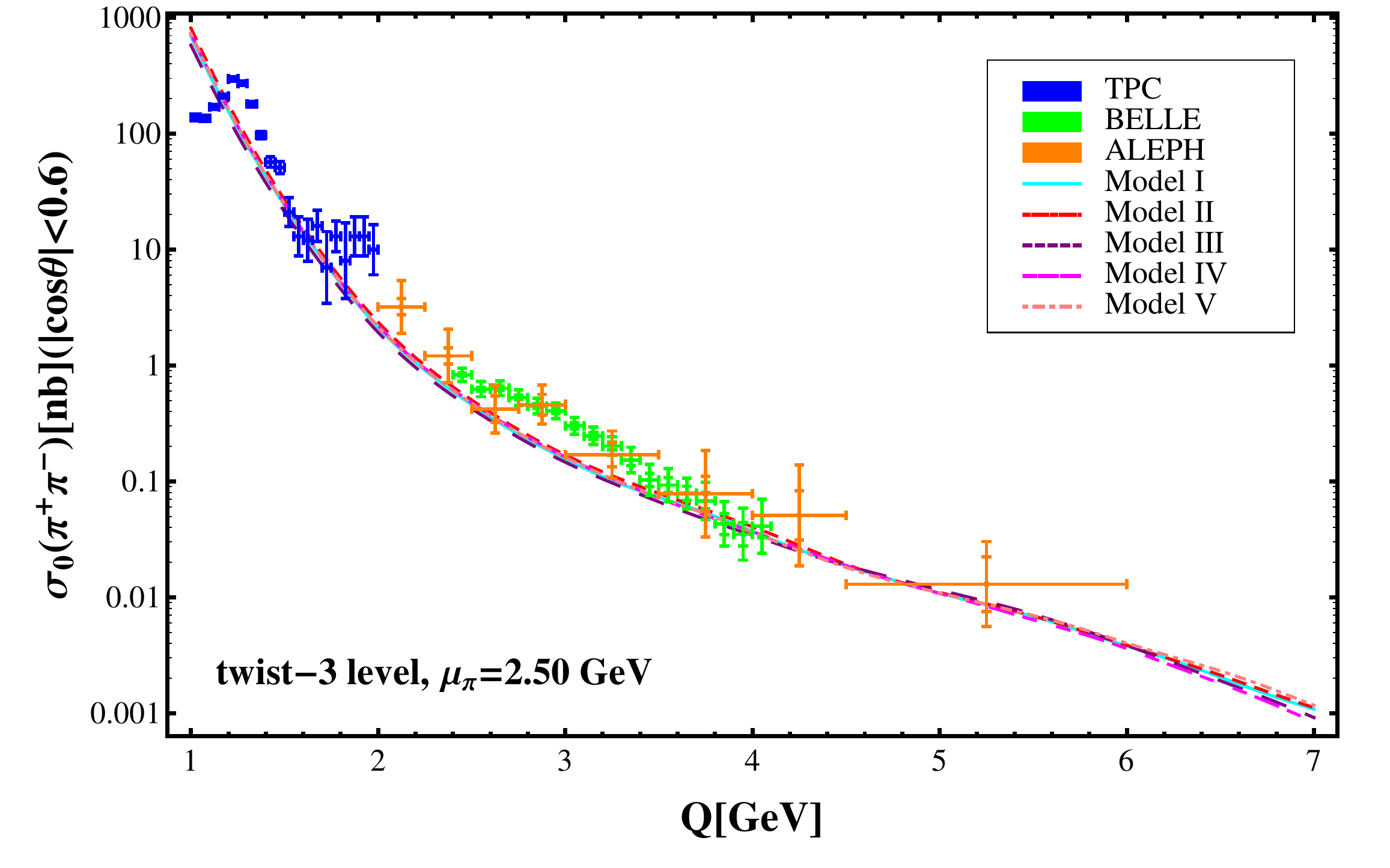}\hspace{0.1cm}
\includegraphics[width=0.32\textwidth]{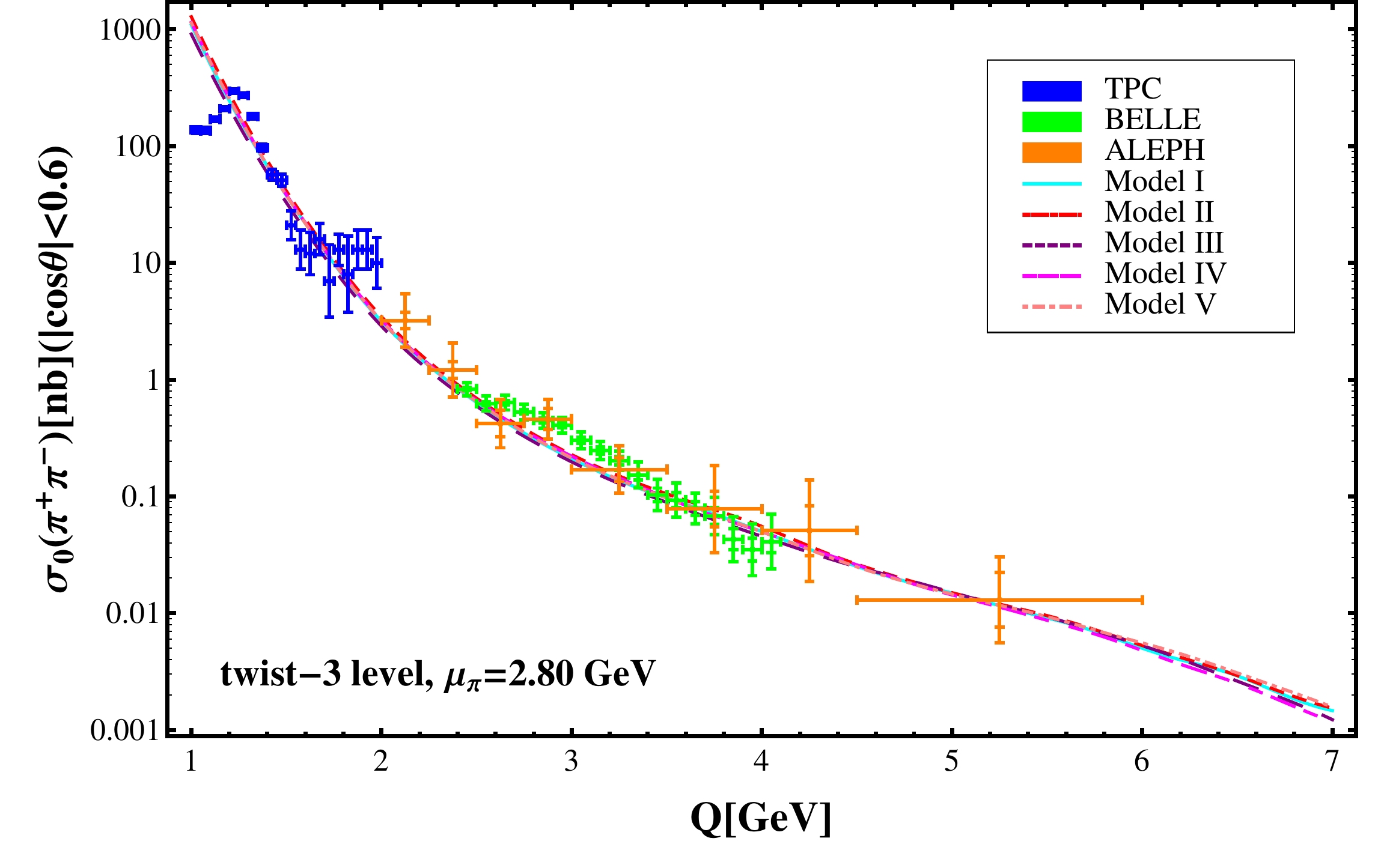}\\
\includegraphics[width=0.32\textwidth]{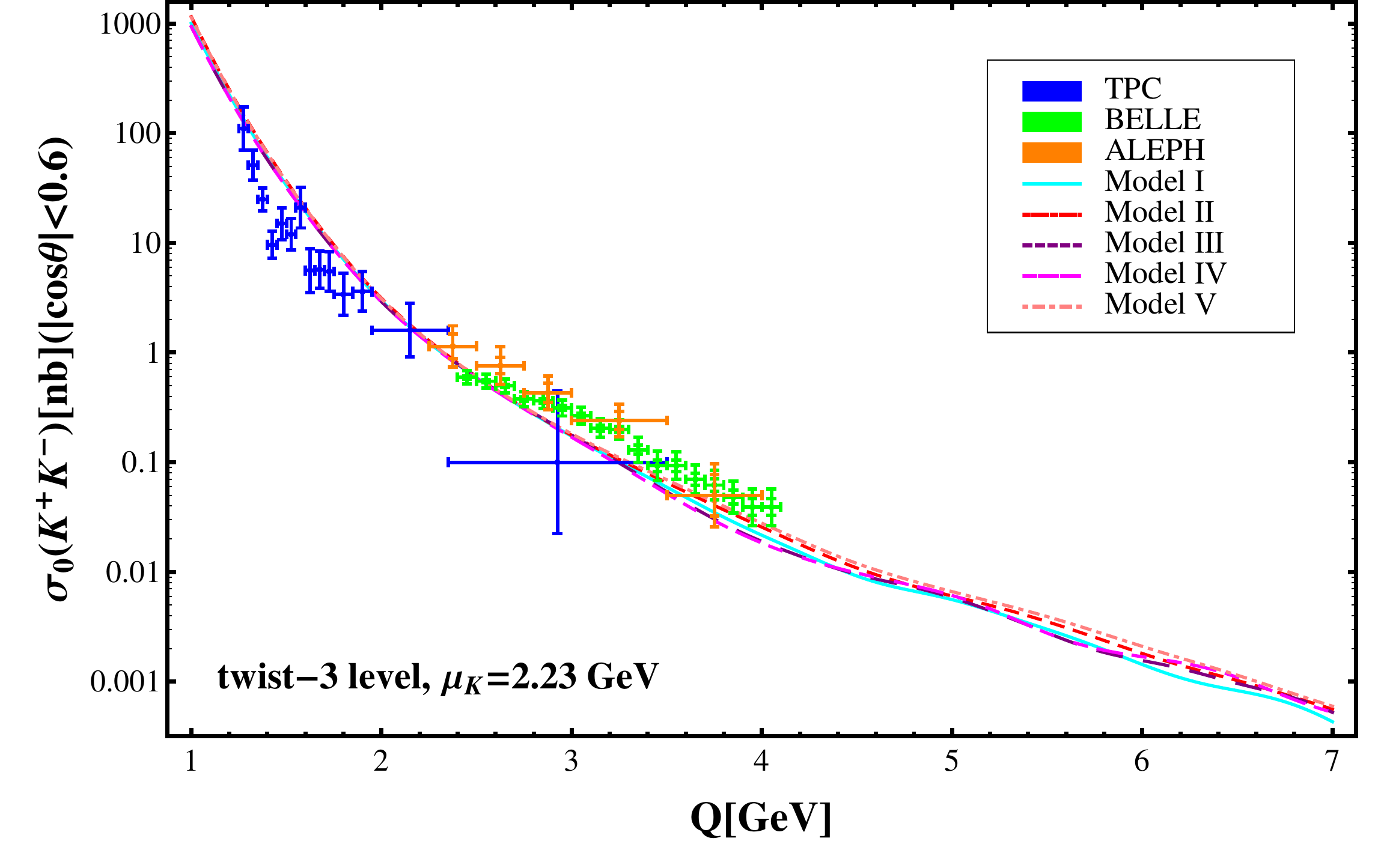}\hspace{0.1cm}
\includegraphics[width=0.32\textwidth]{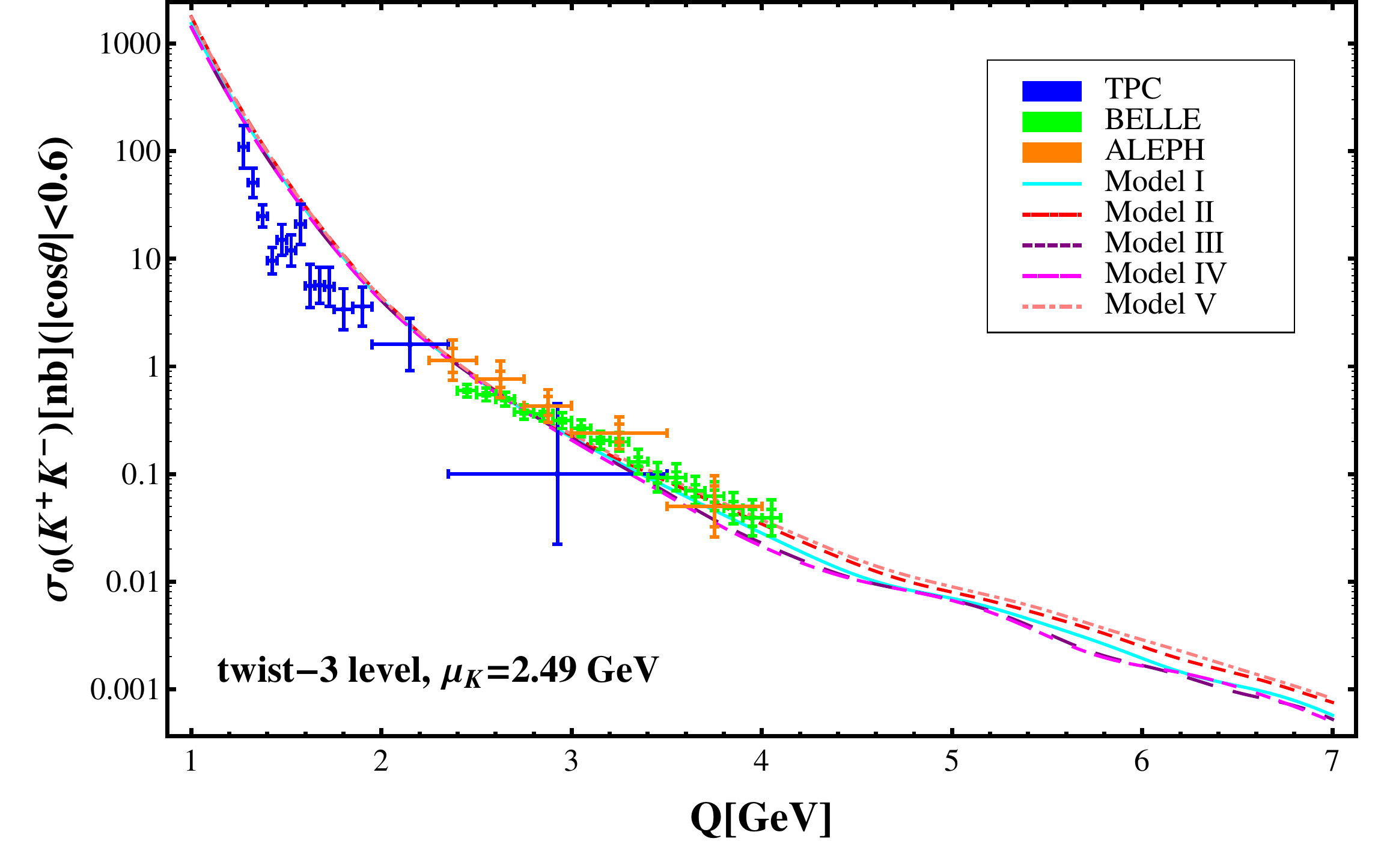}\hspace{0.1cm}
\includegraphics[width=0.32\textwidth]{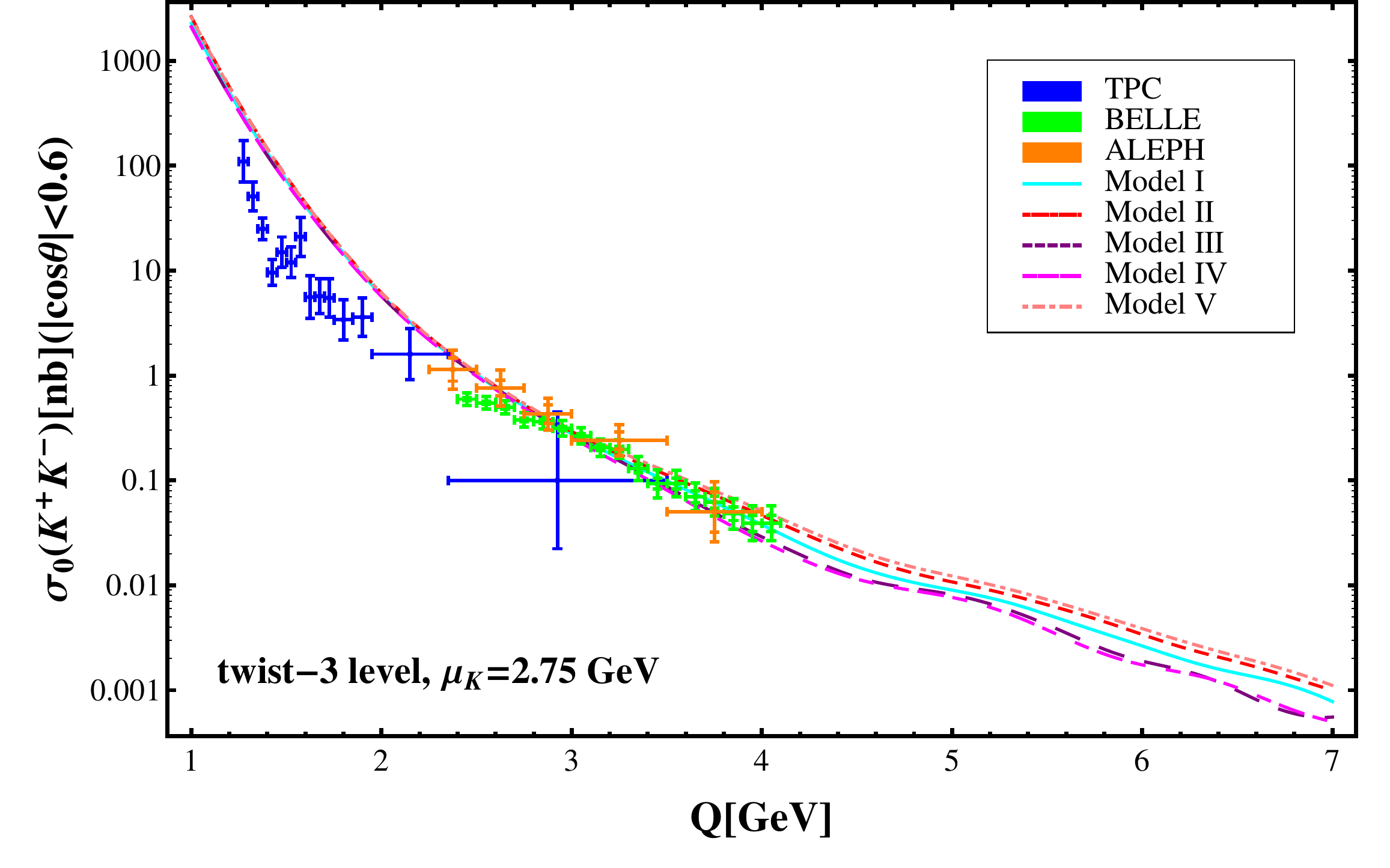}
\caption{Twist-3 results of the cross sections $\sigma_{0}(\pi^{+}\pi^{-})$ and $\sigma_{0}(K^{+}K^{-})$ with the five sample models of the distribution amplitudes listed in Table.~\ref{tab:Geg-moments}. The points with errors are the experimental data from Refs.~\cite{Aihara:1986qk,Heister:2003ae,Nakazawa:2004gu}.}
\label{fig:fig3}
\end{figure}

The theoretical uncertainty in this work mainly comes from the nonperturbative inputs of meson distribution amplitudes: one is the Gegenbauer moments $a_n^M$ and the other is the chiral enhancement parameters $\mu_M$ ($n=2,\ 4$ for pion and $n=1,\ 2$ for kaon, $M=\pi,\ K$). In Table~\ref{tab:Geg-moments}, we present five different models of the distribution amplitudes. And the corresponding Gegenbauer moments are in the region which can cover the ones given in the recent lattice determinations~\cite{Bali:2019dqc}.

With the five models of the pion and kaon distribution amplitudes, the cross sections $\sigma_0(M^{+}M^{-})$ at twist-2 and twist-3 levels are illustrated in Fig.~\ref{fig:fig2} and Fig.~\ref{fig:fig3} respectively.
For the $\pi^{+}\pi^{-}$ cross sections shown in Fig.~\ref{fig:fig2}, one can find that the lines of model I, IV and V with $a_{2}^{\pi}+a_{4}^{\pi}$ close to the same value ($\sim0.23$) almost coincide with each other. And the line of model II (III) with the smallest (largest) $a_{2}^{\pi}$ and $a_{4}^{\pi}$ have the largest (smallest) slope\renewcommand{\thefootnote}{\fnsymbol{footnote}}\footnote[2]{The $Q$-dependence of the cross sections $\sigma_0(M^{+}M^{-})$ can be parameterized into the form: $\sigma_0(M^{+}M^{-})\propto Q^{-n}$~\cite{Nakazawa:2004gu,Duplancic:2006nv}. In this work, we use the ``slope" to indicate the exponent $n$.}. It indicates that the twist-2 result of $\sigma_0(\pi^{+}\pi^{-})$ depends on the Gegenbauer momentums almost in the combination of $a_{2}^{\pi}+a_{4}^{\pi}$. Moreover, by fixing $a_{2}^{\pi}$ ($a_{4}^{\pi}$), i.e., comparing the result of model II with that of model IV (V), the slope of the pion lines is found to become large as $a_{4}^{\pi}$ ($a_{2}^{\pi}$) decrease. The same conclusion can also be obtained by comparing the result of model III with that of model V (IV). While, for the $K^{+}K^{-}$ cross sections shown in Fig.~\ref{fig:fig2}, one can find that the lines of model II and V (III and IV) almost coincide with each other, which implies the twist-2 result of $\sigma_0(K^{+}K^{-})$ is insensitive to the variations of $a_{1}^{K}$. Furthermore, the lines of model II and V (III and IV) with the smallest (largest) $a_{2}^{K}$ have the largest (smallest) slope, i.e., the slope of the kaon lines become large as $a_{2}^{K}$ decrease. However, with all the five distribution amplitudes, we find that the twist-2 result of $\sigma_{0}(K^{+}K^{-})$ is several times smaller than its experiment data~\cite{Aihara:1986qk,Heister:2003ae,Nakazawa:2004gu}, and the twist-2 result of $\sigma_{0}(\pi^{+}\pi^{-})$ is almost an order of magnitude smaller than the data~\cite{Aihara:1986qk,Heister:2003ae,Nakazawa:2004gu}. Similar conclusion was already drawn in Refs.~\cite{Vogt:1999sw,Vogt:2000bz}.

\begin{figure}[!!htb]
\centering
\includegraphics[width=0.47\textwidth]{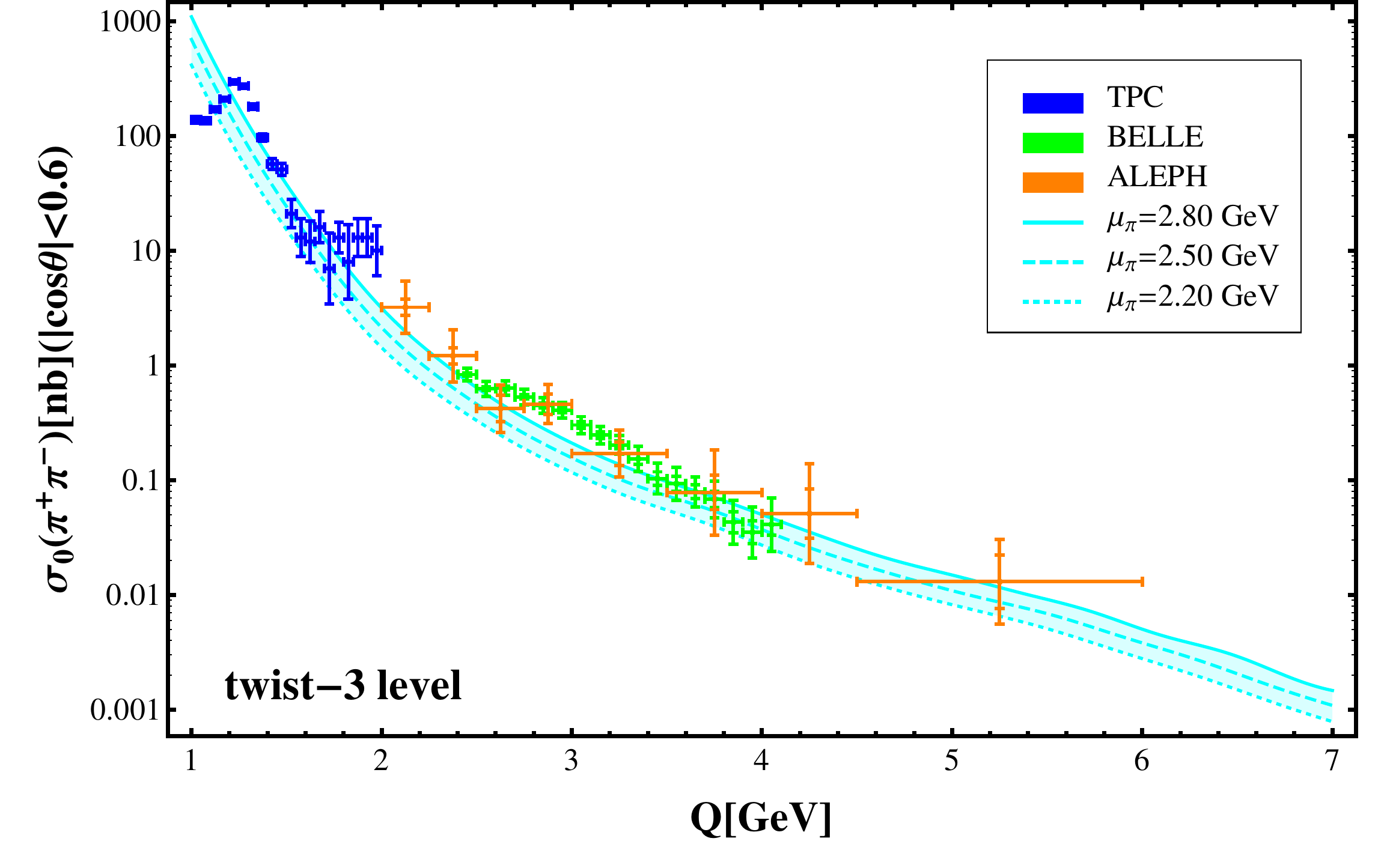}\hspace{0.5cm}
\includegraphics[width=0.47\textwidth]{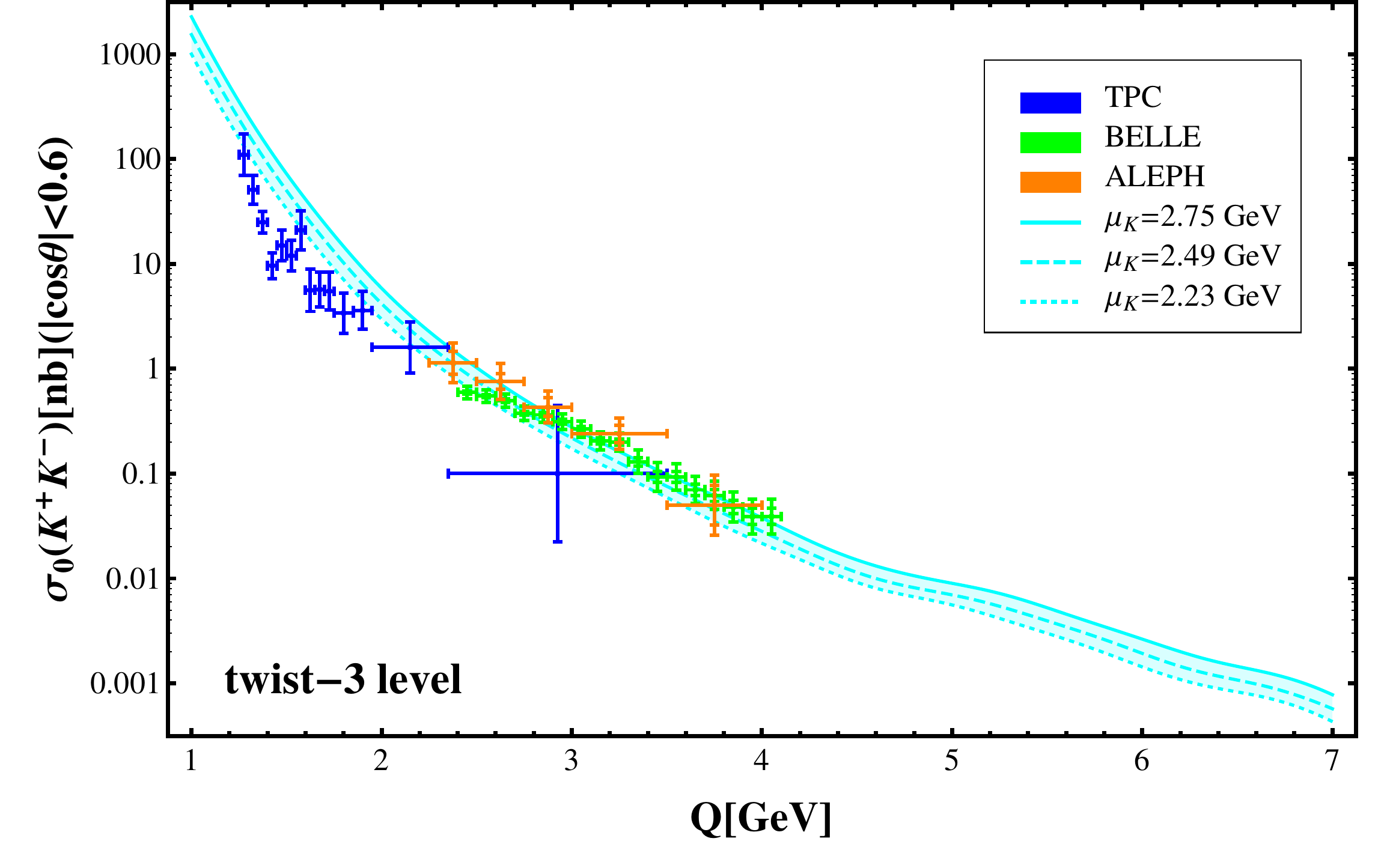}
\caption{Twist-3 results of the cross sections $\sigma_{0}(\pi^{+}\pi^{-})$ and $\sigma_{0}(K^{+}K^{-})$ with uncertainties arising from the chiral enhancement parameters $\mu_{M}$.}
\label{fig:fig4}
\end{figure}

\begin{figure}[!!htb]
\centering
\includegraphics[width=0.47\textwidth]{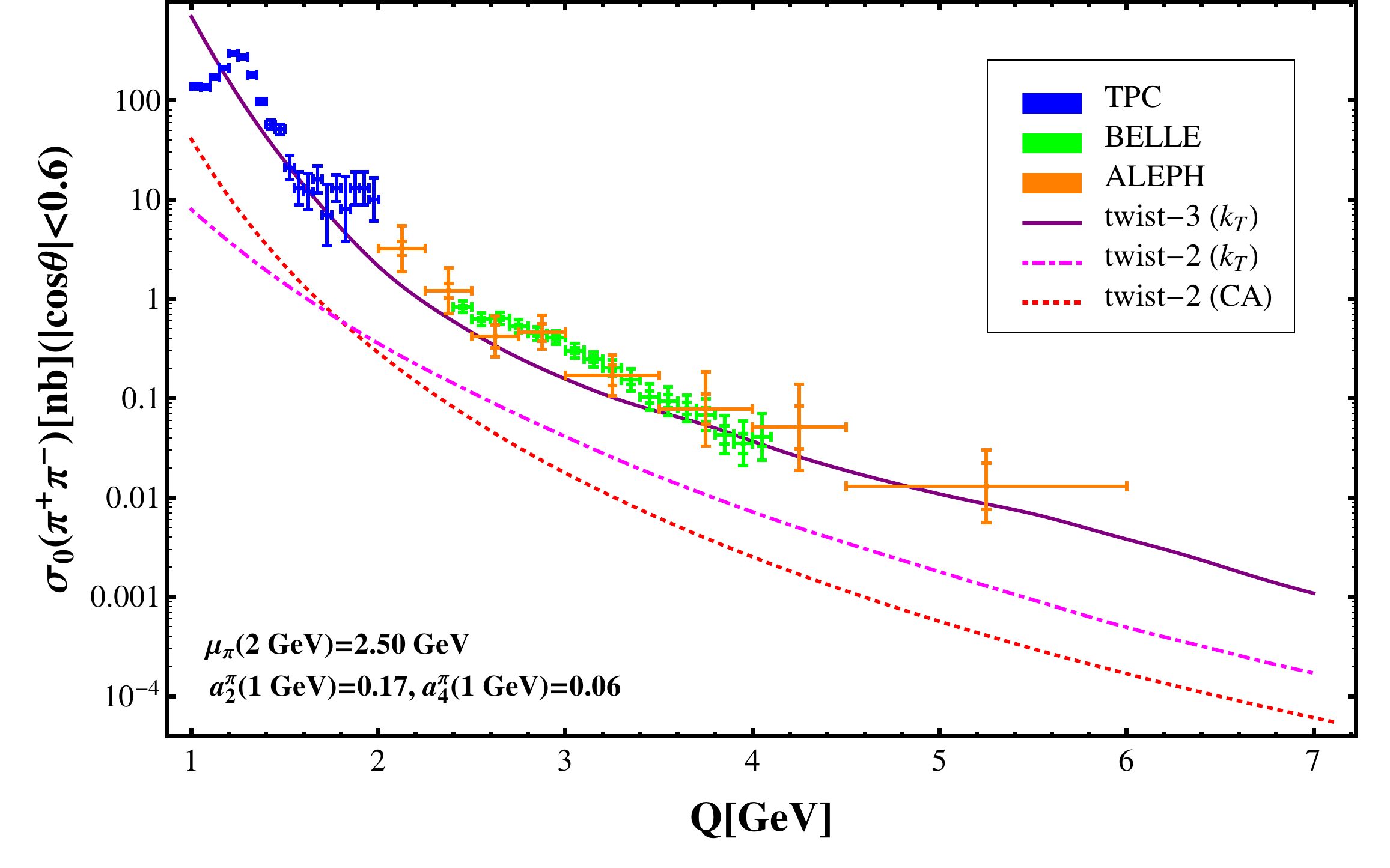}\hspace{0.5cm}
\includegraphics[width=0.47\textwidth]{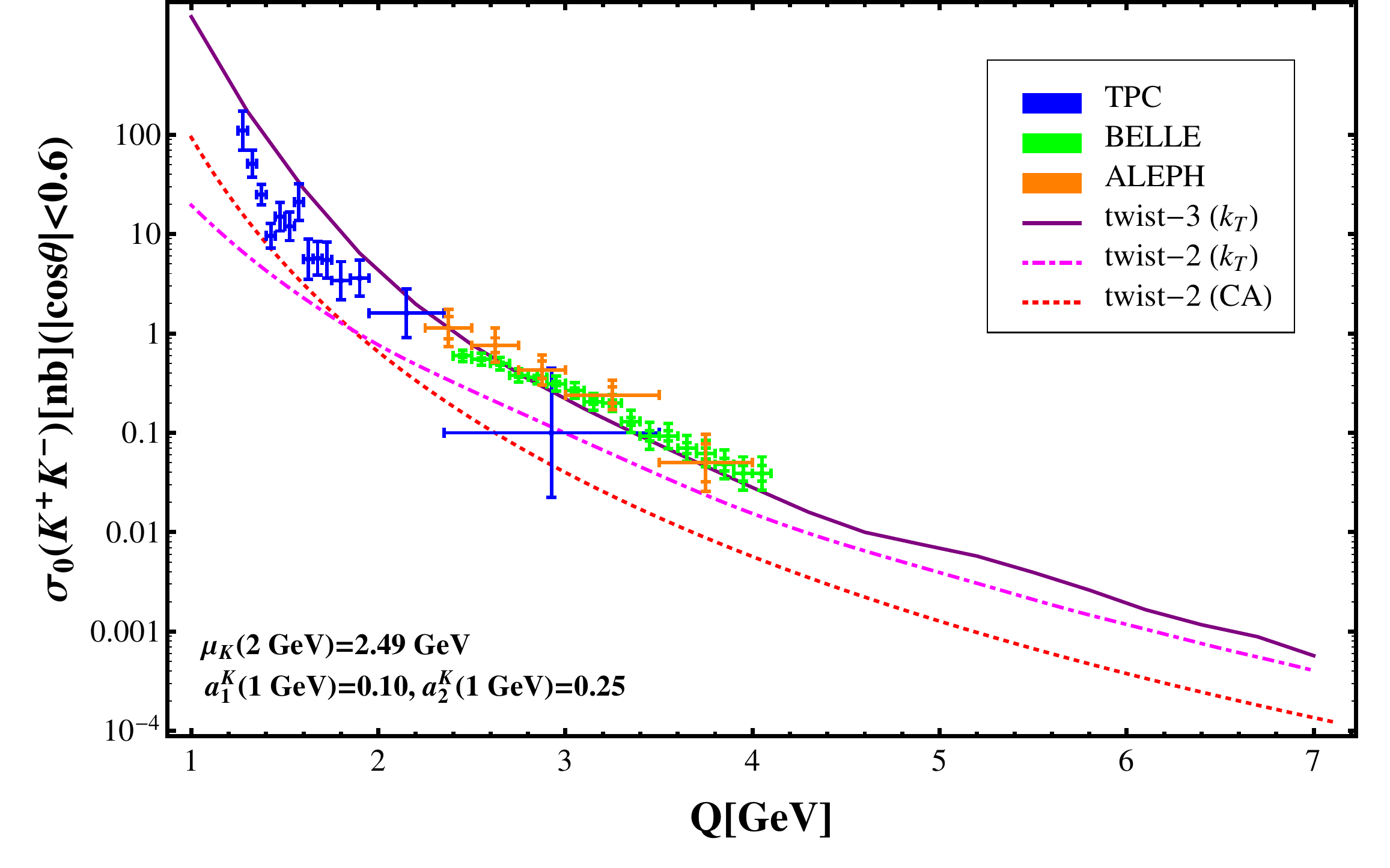}
\caption{Cross sections for $\gamma\gamma\rightarrow\pi^+\pi^-$ and $\gamma\gamma\rightarrow K^+K^-$. Red dotted lines: the twist-2 results in the collinear factorization; Magenta dot-dashed lines: the twist-2 results in the $k_T$ factorization; Purple solid lines: the twist-3 results in the $k_T$ factorization. The points with errors are the experimental data from TPC~\cite{Aihara:1986qk}, BELLE~\cite{Nakazawa:2004gu} and ALEPH~\cite{Heister:2003ae}.}
\label{fig:fig5}
\end{figure}

At twist-3 level, to assess the uncertainties related to the Gegenbauer moments, we draw the figures at specific chiral enhancement parameters. From Fig.~\ref{fig:fig2} and Fig.~\ref{fig:fig3}, one can see that the relative uncertainties come from the Gegenbauer moments in the twist-3 results are much smaller than that in the twist-2 results.
With the Gegenbauer moments chosen at their central values and the chiral enhancement parameters being in their reasonable ranges, the twist-3 cross sections are reshown in Fig.~\ref{fig:fig4}. Comparing Fig.~\ref{fig:fig3} with Fig.~\ref{fig:fig4}, it is found that the main uncertainties in the twist-3 contributions come from the chiral enhancement parameters. For both the $\pi^{+}\pi^{-}$ and $K^{+}K^{-}$ processes, the twist-3 cross sections are found to be in good agreement with the BELLE  measurements~\cite{Nakazawa:2004gu} in the energy region $2.4~\mathrm{GeV}<Q<4.1~\mathrm{GeV}$ and the ALEPH measurements~\cite{Heister:2003ae} in the energy region $2.0~\mathrm{GeV}<Q<6.0~\mathrm{GeV}$. Especially, when the chiral enhancement parameters are taken the upper value $\mu_{\pi}(2\ \mathrm{GeV})=2.80~\mathrm{GeV}$ and $\mu_{K}(2\ \mathrm{GeV})=2.75~\mathrm{GeV}$, the agreements are remarkable. But, in the relatively low energy region $1.0~\mathrm{GeV}<Q<2.0~\mathrm{GeV}$, there still exists considerable discrepancies between our results and the experimental data~\cite{Aihara:1986qk}. For the $\pi^{+}\pi^{-}$ process, the discrepancy in $1.0~\mathrm{GeV}<Q<2.0~\mathrm{GeV}$ may be ascribed to the interference of the continuum with the resonances, such as the $f_{2}(1270)$. For the $K^{+}K^{-}$ process, the neglection of the twist-4 contributions, which are suppressed by the factor $m_{K}^{2}/Q^{2}$, may lead to the discrepancy in $1.0~\mathrm{GeV}<Q<2.0~\mathrm{GeV}$.

In Fig.~\ref{fig:fig5}, we make a comparison between the twist-2 and the twist-3 results of the cross sections $\sigma_{0}(\pi^{+}\pi^{-})$ and $\sigma_{0}(K^{+}K^{-})$ with all the nonperturbative input parameters chosen at the central values. Compared with the previous twist-2 results~\cite{Wang:2015mod} obtained in the collinear factorization, the twist-2 results in the $k_T$ factorization are enhanced several times in the energy region $2.0~\mathrm{GeV}<Q<7.0~\mathrm{GeV}$, but suppressed significantly in the relatively low energy region $1.0~\mathrm{GeV}<Q<2.0~\mathrm{GeV}$. The enhancement in $2.0~\mathrm{GeV}<Q<7.0~\mathrm{GeV}$ may be due to the higher-power corrections from the transverse momentum $\mathbf{k}_{\perp}$. While the suppression in $1.0~\mathrm{GeV}<Q<2.0~\mathrm{GeV}$ may be ascribed to the corrections from the sudakov resummation, which have been organized into the suppression factor $\mathrm{exp}[-S]$. As pointed out in Ref.~\cite{Farrar:1989wb}, the asymptotic scattering amplitudes for the processes $\gamma\gamma\rightarrow\pi^{+}\pi^{-},~K^{+}K^{-}$ are insensitive to the Sudakov corrections. It indicates that the Sudakov factor $\mathrm{exp}[-S]$ affects the cross sections slightly in the relatively high energy region, but may lead to a strong suppression at the relatively low energy scale. Comparing the twist-2 results with the twist-3 ones in the $k_{T}$ factorization, one can conclude that the twist-3 corrections and their interference with the $\mathbf{k}_{\perp}$ effects are very significant for both the $\pi^{+}\pi^{-}$ and $K^{+}K^{-}$ processes in the intermediate energy region $1.0~\mathrm{GeV}<Q<7.0~\mathrm{GeV}$. Especially in the relatively low energy region $1.0~\mathrm{GeV}<Q<2.0~\mathrm{GeV}$, the twist-3 results of $\sigma_{0}(\pi^+\pi^-)$ and $\sigma_{0}(K^+K^-)$ are enhanced by almost an order of magnitude as compared with the twist-2 results. Similar situations can also be found in the calculations of the pion-photon transition form factor~\cite{Hu:2012cp} and the pion electromagnetic form factor~\cite{Raha:2008ve,Li:2010nn,Cheng:2015qra}. A more detailed investigation has shown that the twist-3 corrections fall off rapidly with increasing $Q$ and, beyond $10~\mathrm{GeV}$ region for both pion and kaon, fall below the twist-2 contributions. As is known that, the twist-2 contributions are expected to dominate in the cross section at the asymptotically large momentum transfers. However, as evident from Fig.~\ref{fig:fig5}, the higher twist contributions should be taken into account in the intermediate energy regions $1.0~\mathrm{GeV}<Q<7.0~\mathrm{GeV}$.


\begin{figure}[!!htb]
\centering
\includegraphics[width=0.56\textwidth]{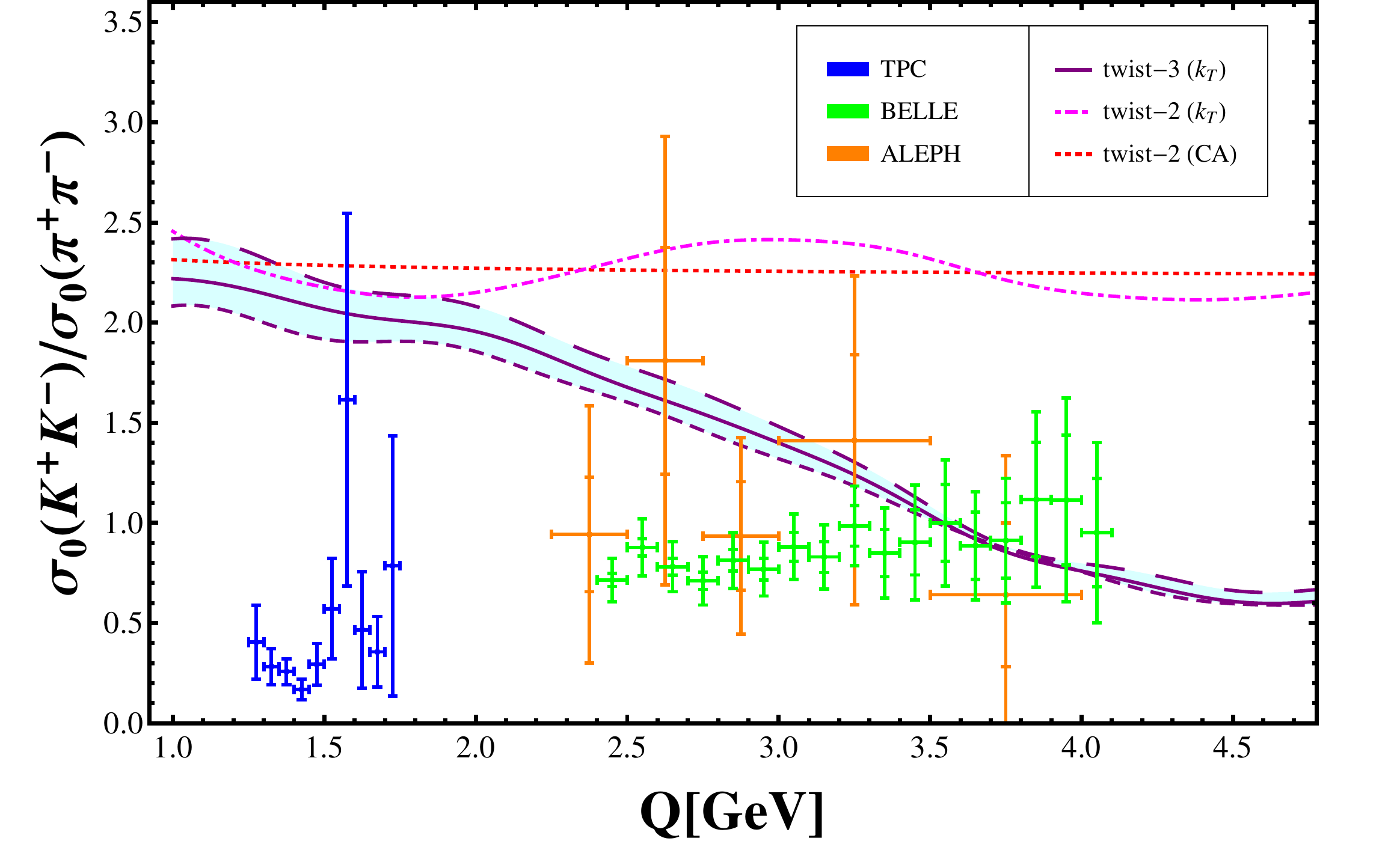}
\caption{\label{fig:fig6}Cross section ratio $\sigma_{0}(K^{+}K^{-})/\sigma_{0}(\pi^{+}\pi^{-})$. Red dotted line: the twist-2 result in the collinear factorization; Magenta dot-dashed line: the twist-2 result in the $k_T$ factorization; Purple lines: the twist-3 result in the $k_T$ factorization. The cyan band indicates the theoretical uncertainty arise from the chiral enhancement parameters $\mu_{M}$. The experimental data are taken from the collaborations of TPC~\cite{Aihara:1986qk}, ALEPH~\cite{Heister:2003ae} and BELLE~\cite{Nakazawa:2004gu}.}
\end{figure}

The cross section ratio $\sigma_{0}(K^{+}K^{-})/\sigma_{0}(\pi^{+}\pi^{-})$ is presented in Fig.~\ref{fig:fig6}. We find that the twist-2 results both in the collinear factorization and in the $k_T$ factorization are almost equal to $(f_{K}/f_{\pi})^{4}\approx 2.26$. Similar theoretical results have also been given in Refs.~\cite{Brodsky:1981rp,Duplancic:2006nv}. While the fitted result from BELLE measurements~\cite{Nakazawa:2004gu} is $0.89\pm0.04\pm0.15$ for $3.0~\mathrm{GeV}<Q<4.1~\mathrm{GeV}$ energy region, which is clearly smaller than the twist-2 results obtained in our work and others~\cite{Brodsky:1981rp,Duplancic:2006nv}. Including the twist-3 corrections, our result of the ratio $\sigma_{0}(K^{+}K^{-})/\sigma_{0}(\pi^{+}\pi^{-})$ is found to be in line with the ALEPH measurements~\cite{Heister:2003ae} in $2.5~\mathrm{GeV}<Q<4.0~\mathrm{GeV}$ and the BELLE measurements~\cite{Nakazawa:2004gu} in $3.5~\mathrm{GeV}<Q<4.1~\mathrm{GeV}$. In the energy region $1.0~\mathrm{GeV}<Q<2.5~\mathrm{GeV}$, there still exists discrepancy between our prediction and the experimental measurements. The reason may be due to the contributions from the resonances and the $m_{M}^{2}/Q^{2}$ terms, which have been neglected in our calculations, may play a significant role in this energy region.


\begin{figure}[!!htb]
\centering
\includegraphics[width=0.35\textwidth]{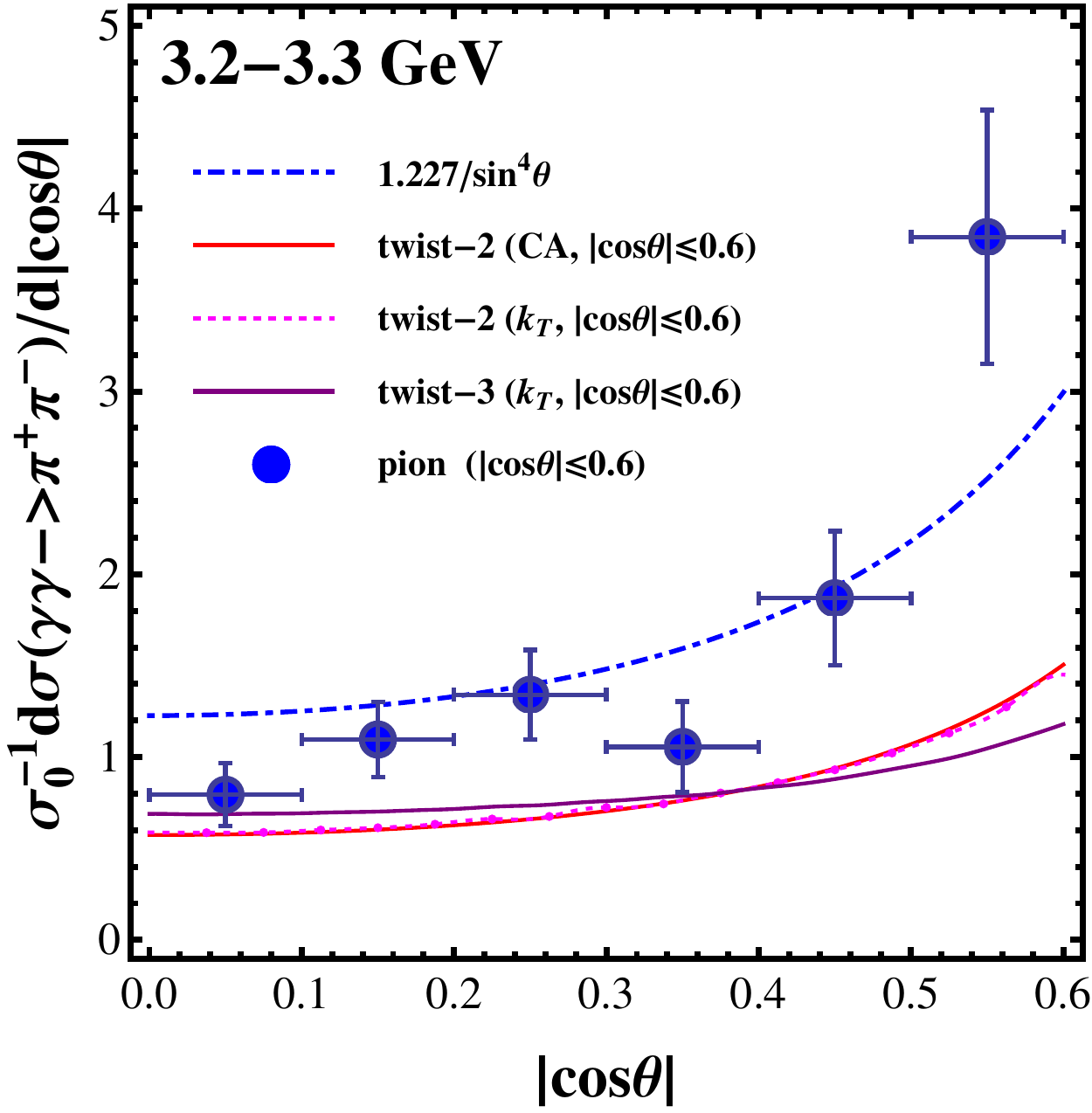}\hspace{0.5cm}
\includegraphics[width=0.35\textwidth]{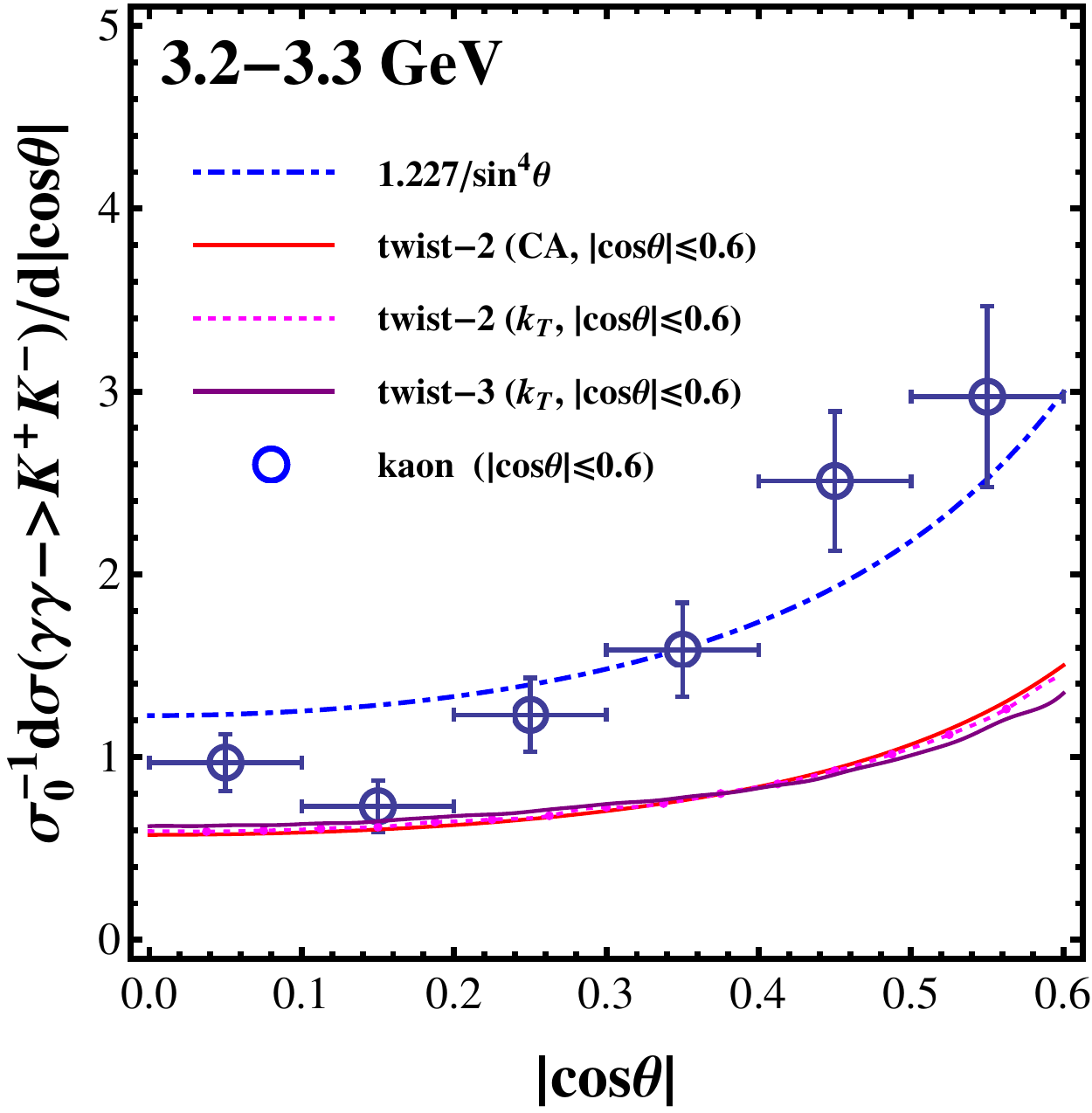}
\caption{Angular distributions $\sigma_{0}^{-1}\mathrm{d}\sigma/\mathrm{d}|\cos\theta|$ for the processes $\gamma\gamma\rightarrow M^{+}M^{-}$. Red solid line: the twist-2 results in the collinear factorization; Magenta dashed line: the twist-2 results in the $k_T$ factorization; Purple solid lines: the twist-3 results in the $k_T$ factorization. All the nonperturbative input parameters are chosen at their central values. The experimental points are from Ref.~\cite{Nakazawa:2004gu}.}
\label{fig:fig7}
\end{figure}

At last but not least, we exhibit the pion and kaon angular distributions, i.e., the differential cross sections $\sigma_{0}^{-1}\mathrm{d}\sigma(\gamma\gamma\rightarrow M^{+}M^{-})/\mathrm{d}|\cos\theta|$, which have been normalized to the cross section $\sigma_{0}(M^{+}M^{-})$. Owing to the normalization, the theoretical uncertainties of the pion and kaon angular distributions from the Gegenbauer moments and the chiral enhancement parameters cancel to a large extent.
Fig.~\ref{fig:fig7} shows our results for $Q=3.2-3.3\ \mathrm{GeV}$ which is in the middle range of the BELLE measurements. One can find the twist-2 results in the collinear factorization and those in the $k_T$ factorization give nearly the same angular distributions. While the angular distributions with the twist-3 corrections in the $k_T$ factorization become more steady over the region $|\cos\theta|<0.6$. Obviously, the angular distributions obtained in this work are not in line with the BELLE measurements~\cite{Nakazawa:2004gu}. The reason may be related to the simplification made in Eq.~(\ref{Thak}), which is illegal in some angular coverage. In Fig.~\ref{fig:fig7}, we also show the known BL result~\cite{Brodsky:1981rp}, corresponding to the blue dot-dashed line $1.227/\sin^{4}\theta$, which is obtained by using the approximate relation in Eq.~(\ref{relation}) with $F_{M}(Q^{2})\approx 0.4~\mathrm{GeV}^{2}/Q^{2}$. However, the CLEO measurements~\cite{Pedlar:2005sj} gave a larger form factor $F_{M}(Q^{2})\approx 1.01~\mathrm{GeV}^{2}/Q^{2}$, which would put the known BL result~\cite{Brodsky:1981rp} in question. As a matter of fact, the approximate relation in Eq.~(\ref{relation}) was obtained at the twist-2 level, and it may not hold in the intermediate energy region, where the twist-3 corrections play a key role in both the scattering processes $\gamma\gamma\rightarrow M^{+}M^{-}$ (from Fig.~\ref{fig:fig5}) and the timelike electromagnetic form factor $F_{M}(Q^{2})$~\cite{Chen:2009sd,Raha:2010kz,Hu:2012cp,Cheng:2019ruz}.

\section{Summary and conclusion}
Our work represents the first investigation of the twist-3 corrections to the two-photon processes $\gamma\gamma\rightarrow\pi^+\pi^-,K^+K^-$ in the perturbative QCD approach based on the $k_{T}$ factorization theorem. The transverse momentum dependence and the resummation effects are included in the perturbative QCD approach. We use the twist-2 and twist-3 light-cone wave functions incorporating transverse degrees of freedom as the nonperturbative dynamical inputs. The nonperturbative contributions from the end-point regions can be effectively suppressed by the Sudakov factor $\mathrm{exp}[-S]$ and the threshold resummation factor $S_t(x)$. Consequently, the perturbative QCD calculation becomes more self-consistent and applicable, especially in the few GeV region.

Within the uncertainties from the distribution amplitudes, the twist-2 results of the cross sections $\sigma_{0}(\pi^{+}\pi^{-})$ and $\sigma_{0}(K^{+}K^{-})$ are much smaller than the experimental data~\cite{Aihara:1986qk,Heister:2003ae,Nakazawa:2004gu}. As shown in our numerical analysis, it is found that both the transverse momentum effects and the twist-3 corrections play a significant role in the two-photon processes $\gamma\gamma\rightarrow\pi^+\pi^-,K^+K^-$ in the intermediate energy region.
In the relatively high energy regions, the twist-3 results of the cross sections
$\sigma_{0}(\pi^{+}\pi^{-})$, $\sigma_{0}(K^{+}K^{-})$ and their ratio $\sigma_{0}(K^{+}K^{-})/\sigma_{0}(\pi^{+}\pi^{-})$ are in good agreement with their corresponding experiment data~\cite{Heister:2003ae,Nakazawa:2004gu}.
But in the relatively low energy region, the predicted cross sections and their ratio are still in disagreement with the experimental measurement~\cite{Aihara:1986qk}. From the above analysis, we can conclude that the cross sections for $\gamma\gamma\rightarrow\pi^+\pi^-,K^+K^-$ with angular region $|\cos\theta|<0.6$ may be dominated by the perturbative QCD contributions when the collision energy $Q\gtrsim 2~\mathrm{GeV}$. And it is noteworthy that, by analyzing the differential cross sections, the authors in Refs.~\cite{Coriano:1994nh,Coriano:1998ge} have concluded that the transition from nonperturbative to perturbative QCD in the processes $\gamma\gamma\rightarrow\pi^+\pi^-,K^+K^-$ is at $Q\approx 2~\mathrm{GeV}$ and $\theta\approx40^{\circ}$. Obviously, these two conclusions are consistent with each other.

However, our results of the pion and kaon angular distributions are different from the BELLE measurements~\cite{Nakazawa:2004gu}. The reason may be partly due to the simplification made in Eq.~(\ref{Thak}), which is illegal in some angular coverage. In addition, in the relatively low energy region $1.0~\mathrm{GeV}<Q<2.0~\mathrm{GeV}$, two-photon processes $\gamma\gamma\rightarrow\pi^{+}\pi^{-},~K^{+}K^{-}$ have been investigated by other methods, such as the QCD sum rule~\cite{Coriano:1994nh,Coriano:1994sz} and the chiral perturbation theory~\cite{Klevansky:2016abt,Hoferichter:2019nlq},
which can be seen as the complementation of the perturbative QCD approach for a full understanding of these processes.

\bigskip
\begin{acknowledgments}
C.~Wang would like to thank Ya-Dong~Yang for valuable discussions and the English language revision. This work is supported by the National Natural Science Foundation of China under Grant Nos.~11675061, 11775092 and 11435003.
\end{acknowledgments}

\newpage
\appendix
\section{The Sudakov function $s(\xi,b,Q)$}
The expression of the Sudakov function $s(\xi,b,Q)$ appearing in Eq.~(\ref{SE}) has the form~\cite{Botts:1989kf,Li:1992nu,Dahm:1995ne}:
\begin{align}\label{SF}
s(\xi,b,Q)=&\dfrac{A^{(1)}}{2\beta_{0}}\hat{q}\mathrm{ln}\left(\dfrac{\hat{q}}{\hat{b}}\right)+\dfrac{A^{(2)}}{4\beta_{0}^2}\left(\dfrac{\hat{q}}{\hat{b}}-1\right)
-\dfrac{A^{(1)}}{2\beta_{0}}\left(\hat{q}-\hat{b}\right)-\dfrac{A^{(1)}\beta_{1}}{4\beta_{0}^3}\hat{q}\left[\frac{\mathrm{ln}(2\hat{b})+1}{\hat{b}}-
\frac{\mathrm{ln}(2\hat{q})+1}{\hat{q}}\right]\notag \\
&-\left[\dfrac{A^{(2)}}{4\beta_{0}^2}-\dfrac{A^{(1)}}{4\beta_{0}}\mathrm{ln}\left(\dfrac{e^{2\gamma-1}}{2}\right)\right]\mathrm{ln}\left(\frac{\hat{q}}{\hat{b}}\right)
+\dfrac{A^{(1)}\beta_{1}}{8\beta_{0}^3}\left[\mathrm{ln}^2(2\hat{q})-\mathrm{ln}^2(2\hat{b})\right]
\end{align}
with the variables
\begin{equation}
\hat{q}\equiv\mathrm{ln}\left(\dfrac{\xi Q}{\sqrt{2}\Lambda_{QCD}}\right),\quad\hat{b}\equiv\mathrm{ln}\left(\dfrac{1}{b\Lambda_{QCD}}\right),
\end{equation}
and the coefficients $A^{(i)}$ and $\beta_i$
\begin{gather}
\beta_{0}=\dfrac{33-2n_f}{12},\quad\beta_{1}=\dfrac{153-19n_f}{24},\notag \\
A^{(1)}=\dfrac{4}{3},\quad A^{(2)}=\dfrac{67}{9}-\dfrac{\pi^2}{3}-\dfrac{10}{27}n_f+\dfrac{8}{3}\beta_{0}\mathrm{ln}
\left(\dfrac{e^\gamma}{2}\right).
\end{gather}
Here the number of quark flavors is $n_f=4$ and $\gamma$ is the Euler constant.

On the physical picture, the longitudinal momentum should be larger than the transverse momentum. It means that the function $s(\xi,b,Q)$ is defined for $\hat{q}>\hat{b}$ (i.e., $\xi Q/\sqrt{2}>1/b$) and set to zero for $\hat{q}\leq\hat{b}$. The range of validity of Eq.~(\ref{SF}) for the Sudakov function is limited to not too small values for the transverse separation of quark and antiquark in the meson. Whenever $b\leq \sqrt{2}/\xi Q$ (i.e., $\hat{b}\geq\hat{q}$), the gluonic corrections are considered as higher-order corrections and absorbed in the hard scattering amplitude, hence they are not contained in the Sudakov factor. Moreover, the complete Sudakov factor $\mathrm{exp}{[-S]}$ is set to unity, if $\mathrm{exp}{[-S]}>1$, which is the case in the small $b$ region. As $b$ increases the Sudakov factor $\mathrm{exp}{[-S]}$ decreases, reaching zero at $b=1/\Lambda_{QCD}$. For $b$ larger than $1/\Lambda_{QCD}$, the true soft region, the Sudakov factor is zero.

\newpage

%

\end{document}